\documentclass{ieeeaccess}
\usepackage{cite}
\usepackage{printlen}
\usepackage{amsmath,amssymb,amsfonts}
\usepackage{mathtools}
\usepackage{cuted}
\usepackage{graphicx}
\usepackage{textcomp}
\def\BibTeX{{\rm B\kern-.05em{\sc i\kern-.025em b}\kern-.08em
    T\kern-.1667em\lower.7ex\hbox{E}\kern-.125emX}}
\usepackage[noend]{algpseudocode}
\usepackage{flushend}
\usepackage{dcolumn} 
\usepackage{bm} 
\usepackage{braket}
\usepackage{xspace}
\usepackage{booktabs}
\usepackage{caption}
\usepackage{ulem}
\usepackage{hyperref} 
\hypersetup{
    colorlinks = true,
    citecolor = magenta,
    linkcolor = purple
}

\usepackage[svgnames, table]{xcolor}
\usepackage{tabularx, makecell, linegoal}

\usepackage{ifthen}

\newboolean{ShowComments}
\setboolean{ShowComments}{false}  
\ifthenelse{\boolean{ShowComments}}%
	{
		\newcommand{\ColorComment}[3]{%
				{\colorbox{#1}{\color{White}   \textsf{\textbf{#2}}} \textcolor{#1}{#3}}}

	}%
	{
		\newcommand{\ColorComment}[3]{}

	}%

\definecolor{rdvcolor}{rgb}{0,0.5,0}
\definecolor{michalcolor}{RGB}{255,127,80}
\definecolor{naphancolor}{RGB}{165,6,245}
\definecolor{joaquincolor}{rgb}{0,0.5,0.8}
\definecolor{robertcolor}{rgb}{0,0.5,0}
\definecolor{allencolor}{RGB}{254,0,0}
\definecolor{alexcolor}{RGB}{75,44,249}

\newcommand{\quisp}{QuISP\xspace}
\newcommand{\sequence}{SeQUeNCe\xspace}

\begin{document}
\history{Date of publication xxxx 00, 0000, date of current version xxxx 00, 0000.}
\doi{10.1109/TQE.2020.DOI}

\title{Cross-Validation of Open-Source Quantum Network Simulators}
\author{
\uppercase{Joaquin Chung}\authorrefmark{1,2,*} \IEEEmembership{Senior Member, IEEE},
\uppercase{Michal Hajdu\v{s}ek}\authorrefmark{3,*},\\
\uppercase{Naphan Benchasattabuse}\authorrefmark{3},
\uppercase{Robert J. Hayek}\authorrefmark{1,4},
\uppercase{Alexander Kolar}\authorrefmark{1,2},\\
\uppercase{Ansh Singal}\authorrefmark{1,4},
\uppercase{Kento Samuel Soon}\authorrefmark{5},
\uppercase{Kentaro Teramoto}\authorrefmark{6},\\
\uppercase{Allen Zang}\authorrefmark{1,2},
\uppercase{Caitao Zhan}\authorrefmark{1},
\uppercase{Raj Kettimuthu}\authorrefmark{1,2} \IEEEmembership{Senior Member, IEEE}, and
\uppercase{Rodney Van Meter}\authorrefmark{7} \IEEEmembership{Senior Member, IEEE}}

\address[1]{Data Science and Learning Division, Argonne National Laboratory, 9700 S. Cass Ave., Lemont, IL 60439, USA}
\address[2]{Pritzker School of Molecular Engineering, University of Chicago, 5801 S. Ellis Ave., Chicago, IL 60637, USA}
\address[3]{Graduate School of Media Design, Keio University, Kanagawa, Japan}
\address[4]{Electrical and Computer Engineering, Northwestern University, 633 Clark St., Evanston, IL 60208, USA}
\address[5]{Graduate School of Media and Governance, Keio University, Kanagawa, Japan}
\address[6]{Mercari, Inc., Minato, Tokyo, Japan}
\address[7]{Faculty of Environment and Information Studies, Keio University, Kanagawa, Japan}
\address[*]{Chung and Hajdu\v{s}ek contributed equally to this paper.}

\tfootnote{
This material is based upon work supported by the U.S.~Department of Energy Office of Science National Quantum Information Science Research Centers as part of the Q-NEXT Center.
This work was supported by the JST Moonshot R\&D program under Grant Numbers JPMJMS226C and JPMJMS256K.
}

\markboth
{Chung, Hajdu\v{s}ek \headeretal: Cross-Validation of Open-Source Quantum Network Simulators}
{Chung, Hajdu\v{s}ek \headeretal: Cross-Validation of Open-Source Quantum Network Simulators}

\corresp{Corresponding authors: Joaquin Chung (email: chungmiranda@anl.gov) and Michal Hajdu\v{s}ek (email: michal@sfc.wide.ad.jp).\\
An earlier version of this paper was presented in part at the IEEE INFOCOM 2025 - IEEE Conference on Computer Communications Workshops (INFOCOM WKSHPS) [DOI: 10.1109/INFOCOMWKSHPS65812.2025.11152951].}

\begin{abstract}
We present a comparative analysis of two open-source quantum network simulators, QuISP and SeQUeNCe, focusing on cross-validation of basic networking tasks to ensure consistency and accuracy in simulation outputs. 
Although both simulators are designed to study quantum network management policies and their dynamics under complex noise models, their differing underlying assumptions can lead to variations in simulation results.
We highlight the discrepancies in how the two simulators handle connections, internal network node processing time, and classical communication, resulting in significant quantitative and qualitative differences in the time required to perform basic network tasks such as entanglement generation, purification, and swapping.
We simulate common scenarios between the simulators to compare both the latency of resource distribution and the fidelity of the distributed resources.
Our findings indicate that while the simulators differ in the time required to complete network tasks---a constant factor difference attributable to their respective connection models---they agree on the fidelity of the distributed resources under identical error parameters.
This work demonstrates a crucial first step towards enhancing the reliability and reproducibility of quantum network simulations, as well as leading to full protocol development.
Furthermore, our benchmarking methodology establishes a foundational set of tasks for the cross-validation of simulators to study future quantum networks.
\end{abstract}

\begin{keywords}
Quantum networks, quantum network simulators, QuISP, SeQUeNCe, discrete event simulation, open-source, validation
\end{keywords}

\titlepgskip=-15pt

\maketitle

\section{Introduction} \label{Introduction}
Quantum networks~\cite{van2014quantum} are an important part of the ongoing second quantum revolution~\cite{dowling2003quantum}.
They hold great promise for enhancing existing functionality of current classical networks (e.g., quantum key distribution~\cite{ekert1991quantum}, enhanced interferometric measurements for telescopes~\cite{gottesman2012longer}, and high-precision clock synchronization~\cite{komar2014quantum}) as well as enabling new applications (e.g., blind quantum computation~\cite{broadbent2009universal} and distributed quantum computation~\cite{cuomo2020towards})~\cite{wang:RFC9583}.
Quantum networks are widely expected to be a necessary component of scalable, fault-tolerant quantum computers~\cite{simmons2024scalable,sakuma2024qfly,naito2026networkbasedqc}, linking nearby quantum processing units via optical quantum switches~\cite{PRXQuantum.2.017002,koyama2024optimal}.
However, the most ambitious goal is to create a quantum internet~\cite{wehner2018quantum} that will connect quantum computing resources across the globe.

Although quantum networks take inspiration from their classical counterparts, they are fundamentally different.
First, quantum-networked applications rely on the consumption of entangled states, a quintessential resource in quantum communications~\cite{hajdusek2023quantum}.
Unlike the store-and-forward nature of packet-switched networks, quantum networks are responsible for the sustained generation of high-quality entangled states shared between two or more remote network nodes.
Second, quantum networks encode quantum information at the level of single photons, meaning that attenuation in optical fiber or free space becomes a major source of loss of entire transmitted states. 
Compounding this detrimental effect, quantum signals cannot be reliably amplified or copied due to the no-cloning theorem~\cite{Park1970,dieks1982communication,zurek1982single}.

Given the aforementioned design constraints, building large-scale quantum networks becomes extremely challenging.
Thus, intermediate quantum nodes---referred to as quantum repeaters---have been introduced to generate short-range, i.e., link-level entanglement~\cite{briegel1998quantum,azuma2023quantum}.
Generally, the generation of link-level entanglement is a non-deterministic process and therefore requires classical messages to be exchanged between the quantum repeaters heralding a success or failure.
Upon successful generation of link-level entanglement between quantum repeaters, the entanglement swapping operation splices two entanglement links into a long-range entangled state shared between the desired end nodes of the quantum network.
This highlights the fact that generation of remote entanglement is akin to coordinated and distributed computation~\cite{vanmeter2022quantum,kozlowski2023rfc}, where quantum repeaters send and receive a constant stream of quantum and classical messages and execute actions based on both local and non-local conditions.

The study of quantum networks quickly becomes intractable using analytical methods as we scale the number of devices and links connecting them, as well as the realism of the network protocols.
Therefore, we must rely on simulations to test design decisions and reason about non-trivial quantum network topologies. For this end, we view a simulator as one (or more) of three things: an X-ray machine, a time machine, or a telescope. A simulator allows us to look inside systems in ways that are otherwise impossible, to project what systems we hope to build in the future will do, or to examine systems at a larger scale than we can (currently) build. Simulators are invaluable tools that bridge mathematical analysis and implementation; in an ideal world, analysis, simulation, and experiment will agree over a broad enough range of conditions to give us confidence in the predictions of the simulator.

Recent years have seen the introduction of a number of quantum network simulators, each being used in the design of real networks.
Examples include NetSquid~\cite{coopmans2021netsquid}, QuantumSavory~\cite{kimlee2025quantumsavory}, QuISP~\cite{satoh2022quisp}, QuNetSim~\cite{diadamo2012qunetsim}, QuNet~\cite{leone2021qunet}, ReQuSim~\cite{wallnofer2024faithfully},
SeQUeNCe~\cite{sequence}, and most recently
Q2NS~\cite{pearson2026Q2NS}.
Despite overlapping aims and goals, these simulators differ significantly in their assumptions, architecture, design, and implementation.
However, the results of the simulations should agree.
When the simulators produce different results on nominally similar configurations, we consider three possibilities: (a) real differences exist due to design choices in the network, (b) valid but differing simplifications in simulation were made at the physical or protocol level, or (c) bugs in one or possibly both simulators exist.
Case (a) represents the differences we wish to study as we commit our design decisions on the way to real-world implementations.
Case (b) represents an opportunity to refine our understanding of how to most accurately simulate quantum networks for parameters of interest. 
Case (c), of course, should simply be fixed (and indeed this exercise resulted in numerous fixes to both simulators used).

The causes exposed above emphasize the need for cross-validation studies of existing quantum network simulators.
In this paper, we present a comparative study of two quantum network simulators: QuISP and SeQUeNCe.
We chose these two simulators for the following reasons: 1) the objective of the tools is to simulate entanglement distribution networks with special focus on the control protocols; 2) the simulators agree on the network architecture but follow different design and implementation patterns; 3) the simulators are open source; and 4) we are the core developers of the simulators.
By focusing on simulation of fundamental quantum networking primitives (e.g., link-level entanglement generation, entanglement swapping, and purification), we identify the discrepancies in how the simulated protocols handle connections, internal node processing time, and classical communication overhead.

We compare the simulators in terms of the predicted time needed to generate a set number of entangled states between two end nodes as well as the fidelity of these distributed states.
We find that the simulators disagree on the predicted time, differing by a constant factor.
Using the differences in their designs, we explain and correctly quantify this offset.
Furthermore, the simulators agree on the predicted fidelities of the final states, pointing to correct implementation of realistic error models in both simulators.
Finally, the experimental methodology presented here represents an important step in developing a full set of cross-validation techniques applicable across all simulators of quantum networks.

We begin by providing a brief overview of the fundamental quantum networking primitives of link-level entanglement generation, entanglement swapping, and purification in Sec~\ref{Preliminaries}.
We then move to describe the two simulators in detail in Sec.~\ref{Simulators}, where we discuss their respective scopes and software architectures.
In Sec.~\ref{Benchmarking}, we detail the four experiments executed on the simulators used for their cross-validation.
Before running the experiments, we develop a detailed theoretical model for the timing of classical messages and quantum operations in the simulators as well as their error models in Sec.~\ref{Models}.
We present the results of the four simulations in Sec.~\ref{Results}, and conclude with discussion in Sec.~\ref{Discussion}.

\section{Preliminaries} \label{Preliminaries}
\begin{figure*}[h!]
    \centering
    \includegraphics[width=0.95\textwidth]{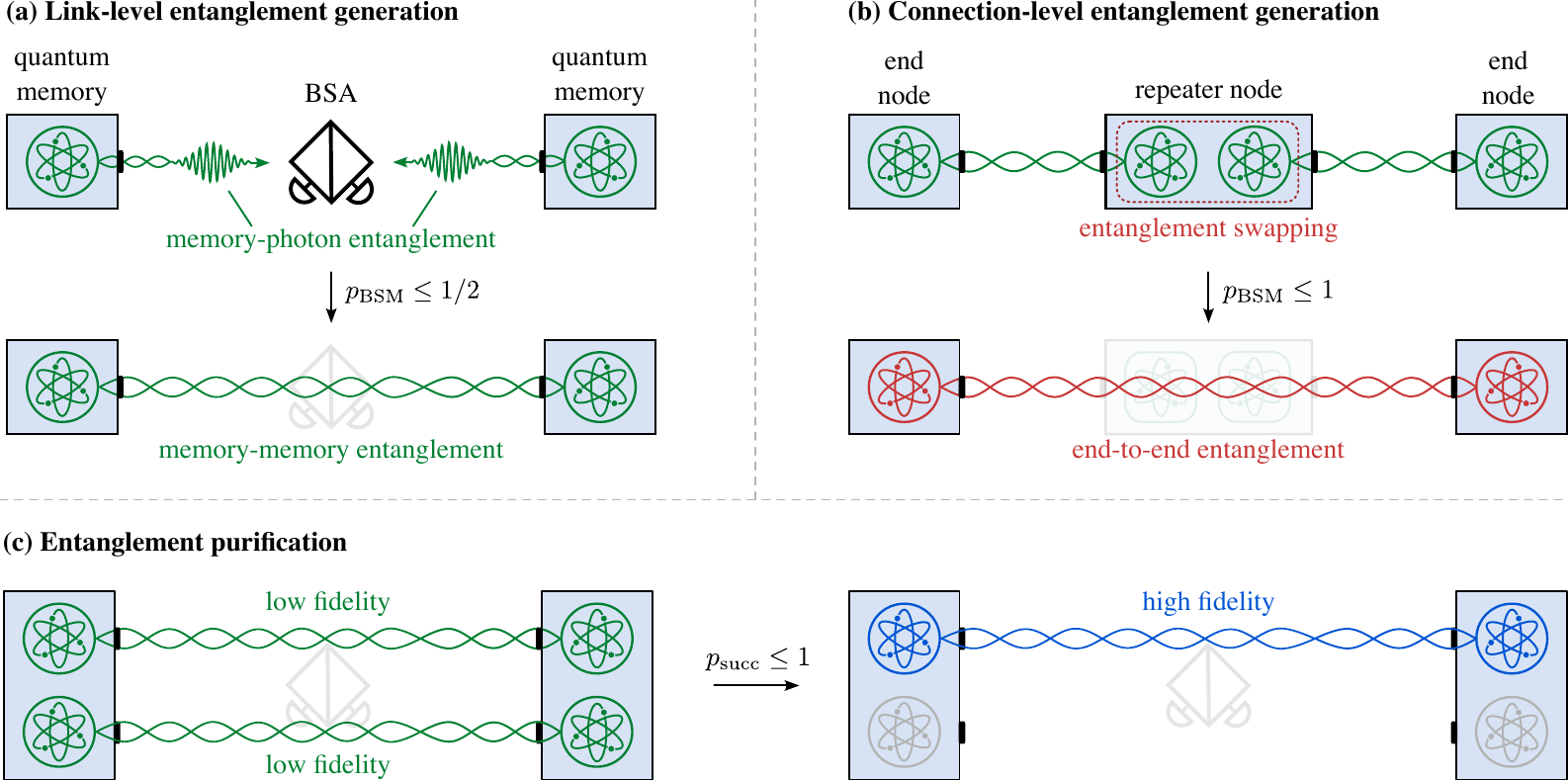}
    \caption{(a) Link-level entanglement generation between neighboring network nodes using a Bell state analyzer (BSA) on photonic qubits emitted by quantum memories. The two photons must arrive at the BSA at the same time, an important task in engineering the protocols and optical path. The probability of success is upper bounded by $1/2$~\cite{lutkenhaus1999bell}, meaning the generation process must be repeated upon failure of the Bell state measurement (BSM). (b) After successful entanglement generation of two neighboring links, the repeater node performs entanglement swapping on its quantum memories to create an end-to-end entangled connection. This process can, in principle, be deterministic. (c) Two link-level entangled states of low fidelity can be transformed into a single entangled state with higher fidelity. This process is probabilistic, and the success probability depends on the initial fidelity of the noisy states.}
    \label{fig:preliminaries}
\end{figure*}

In this section, we describe the basics of entanglement distribution, the central task of quantum networking on which we will base our cross-validation study.
The primary tool used to share entanglement between distant nodes of a quantum network is \textit{entanglement swapping} (ES)~\cite{PhysRevLett.71.4287,jin2015highly,hermans2022qubit}.
We will first outline how ES works before reviewing how it can be applied to photonic qubits to generate entangled states at the link level and, subsequently, to memory-based qubits to generate entanglement at the connection level.

Entanglement is a crucial consumable resource necessary for the distributed quantum applications discussed in the introduction. The job of the quantum network is to create entanglement while managing errors.
Consider the four \textit{Bell pairs}, given by
\begin{equation}
    |\Phi^{\pm}\rangle = \frac{1}{\sqrt{2}}(|00\rangle\pm|11\rangle), \quad |\Psi^{\pm}\rangle = \frac{1}{\sqrt{2}}(|01\rangle\pm|10\rangle).
    \label{eq:Bell_states}
\end{equation}

These two-qubit maximally entangled states are the target states of most entanglement distribution protocols.
Consider the case of two such states, $|\Phi^+\rangle_{AB_1}$ and $|\Phi^+\rangle_{B_2C}$, where the states are shared between network nodes $AB$ and $BC$ with node $B$ being in possession of two qubits.
When performing a \textit{Bell-state measurement} (BSM), node $B$ can project the state of the qubits held by nodes $A$ and $C$ onto one of the Bell pairs in \eqref{eq:Bell_states}. Node $B$ communicates the BSM outcome (i.e., two classical bits) to nodes $A$ and $C$, which then apply the appropriate Pauli corrections to ensure a deterministic distribution of a pre-agreed Bell pair.

One possible way of generating link-level entanglement between neighboring quantum nodes is pictured in Fig.~\ref{fig:preliminaries}(a), and is known as the \textit{memory-interference-memory}, or \textit{meet-in-the-middle} (MIM) link architecture.

The process begins with each quantum memory emitting a photon entangled with its qubit state.
These memory-photon entangled states are the physical representations of the abstract states $|\Phi^+\rangle_{AB_1}$ and $|\Phi^+\rangle_{B_2C}$ above.
The photons are then routed to a \textit{Bell-state analyzer} (BSA), where the photonic (flying) qubits undergo a BSM.
A crucial property of the photonic BSM, when implemented with linear optics, is that the maximum success probability of the measurement is fundamentally limited by $1/2$~\cite{lutkenhaus1999bell}.
This makes link-level entanglement generation an intrinsically probabilistic process, even when all other sources of loss such as attenuation in fiber are disregarded.

After the successful generation of two neighboring entangled links, ES can be applied again, but this time on the memory qubits, as shown in Fig.~\ref{fig:preliminaries}(b).
This transfers the entanglement onto memories located in non-neighboring quantum nodes.
Unlike ES with photonic qubits, BSM performed on memory qubits does not have an upper bound on the success probability, and can therefore be made deterministic with ideal hardware. For either type of ES, after the local BSM operation, the needed Pauli corrections described above will be randomly determined by the BSM output, requiring classical messages to the partners. Following the production of multiple Bell pairs, a purification/distillation protocol that converts multiple low-fidelity pairs into a single high-fidelity Bell pair can be used, as shown in~Fig.~\ref{fig:preliminaries}(c).

\section{The Simulators} \label{Simulators}
In this section we describe the two simulators that are the object of this study: the Quantum Internet Simulation Package (\quisp, see Section~\ref{sec:quisp}) and the Simulator of QUantum Network Communication (\sequence, see Section~\ref{sec:sequence}).
We first provide a brief background about software simulators of classical networks.
Simulators of classical networks have a long history dating back to the 1960s~\cite{gordon1961gpss,real,nest}. 
The most successful ones are based on the discrete event-driven simulation (DES) paradigm.
The DES~\cite{des} approach models the operation of a system as a discrete sequence of events in time.
Each event occurs at a particular instant in time and marks a change of state in the system.
No change in the system is assumed to occur between consecutive events.
The simulator maintains an event queue sorted by the scheduled event execution time. 
The simulation itself is performed by successively processing the events in the queue.
After decades of research and development, a number of open-source network simulation tools have been developed, such as ns-3~\cite{ns3} and OMNeT++~\cite{OMNET}.

Before describing the quantum network simulators in detail, we give a brief overview in terms of design and implementation choices.
We aim to help the reader choose the appropriate tool for their requirements, because as shown in Section~\ref{Results}, both simulators produce similar results.
While QuISP's goal is to simulate large-scale quantum networks and internetworks using minimal resources, SeQUeNCe aims to accurately simulate quantum photonic networks with a high level of customizability.
The current implementation of both simulators focuses on modeling the first generation (1G) quantum repeater networks (i.e., entanglement distribution over a chain of quantum repeaters followed by entanglement swapping without error correction)~\cite{muralidharan2016optimal}.
Table~\ref{tab:sim-comparison} summarizes the main characteristics such as choice of programming language, software architecture, and error models.

\begin{table*}[]
\centering
\caption{Comparison of QuISP and SeQUeNCe design and implementation choices}
\label{tab:sim-comparison}
\resizebox{\textwidth}{!}{%
\begin{tabular}{l l l c l l }
\toprule
 &
  \multicolumn{1}{c}{\textbf{Goal}} &
  \multicolumn{1}{c}{\textbf{Platform}} &
  \textbf{\begin{tabular}[c]{@{}c@{}}Quantum\\ Network\\ Generation~\cite{muralidharan2016optimal}\end{tabular}} &
  \multicolumn{1}{c}{\textbf{Architecture}} &
  \multicolumn{1}{c}{\textbf{Error Models}} \\
  \midrule
\textbf{QuISP} &
  \begin{tabular}[c]{@{}l@{}}To simulate large scale quantum \\ network and internetworks with \\ minimal resources.\end{tabular} &
  \begin{tabular}[c]{@{}l@{}}OMNeT++ and \\ C++\end{tabular} &
  1G &
  \begin{tabular}[c]{@{}l@{}}Each node in the network implements\\ QRSA.\end{tabular} &
  \begin{tabular}[c]{@{}l@{}}Support for qubit error models.\end{tabular} \\
  \midrule
\textbf{SeQUeNCe} &
  \begin{tabular}[c]{@{}l@{}}To accurately simulate photonic \\ quantum networks, while providing \\ high levels of customizability.\end{tabular} &
  Python 3.12+ &
  1G &
  \begin{tabular}[c]{@{}l@{}}Simulation kernel plus applications,\\ hardware, entanglement, network,\\ and resource management modules.\end{tabular} &
  \begin{tabular}[c]{@{}l@{}}Errors are encoded as \\ variable hardware parameters.\end{tabular} \\
  \bottomrule
\end{tabular}%
}
\end{table*}

\subsection{QuISP} \label{sec:quisp}
The Quantum Internet Simulation Package (QuISP)~\cite{satoh2022quisp,quisp-github} is an open-source package designed to simulate large-scale quantum networks and internetworks.
It is built on top of OMNeT++~\cite{varga2010omnetpp}, a discrete event-driven C++ library used for building classical network simulators.
The primary goals of QuISP are to aid in protocol design, architecture and performance prediction, and evaluation of dynamic and emergent behavior of quantum networks. A large emphasis of QuISP's design is placed on enabling the simulation of quantum networks at scale. Its goal is to be able to simulate an internetwork consisting of 100 networks with 100 nodes each, while using minimal computational resources.

There is currently no consensus for the best underlying architecture of quantum networks, though certain common features and principles are slowly beginning to emerge~\cite{vanmeter2022quantum,kozlowski2023rfc}.
QuISP assumes that each node of a quantum network is equipped with one or more \textit{quantum network interface cards} (QNICs) containing several quantum memories, which are used for storage and processing of quantum information.
Inter-node quantum communication uses single photons to establish entangled pairs between nearest-neighbor quantum network nodes.
Currently, QuISP supports three link-level architectures: memory-interference-memory (MIM)~\cite{jones2016design}, heterogeneous links combining BSA and EPS~\cite{soon2024implementation}, and ground-to-satellite links~\cite{fittipaldi2024entanglement}.

\begin{figure*}
    \centering
    \includegraphics[width=0.9\textwidth]{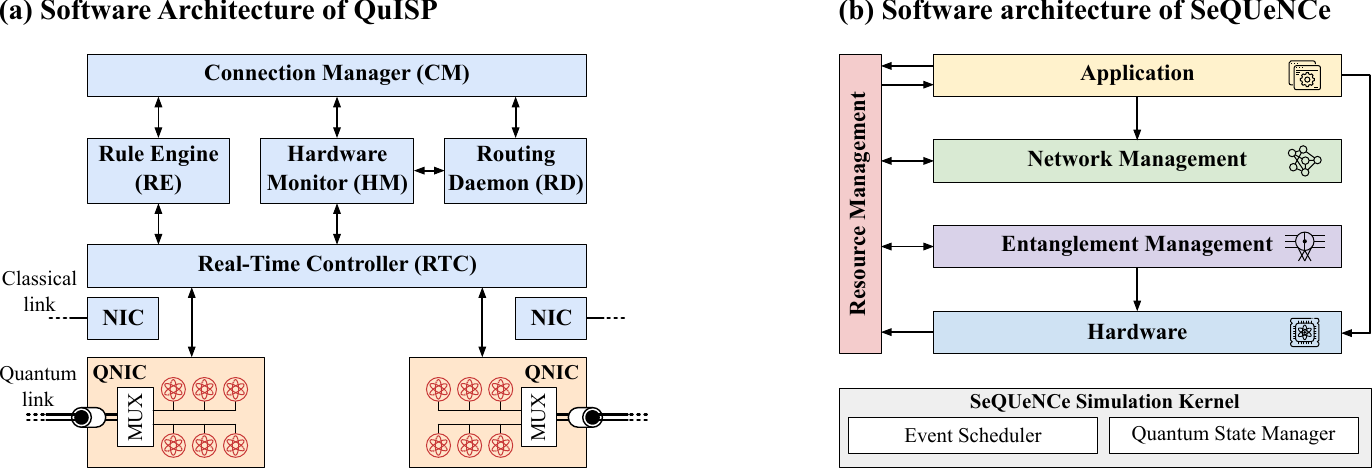}
    \caption{(a) Quantum Router Software Architecture (QRSA) used in \quisp to control the behavior of network nodes. (b) \sequence software framework comprises a simulation kernel and five modules for applications, hardware models, entanglement, network, and resource management.}
\label{fig:software_arch}
\end{figure*}

To control and coordinate the operation of the quantum nodes, QuISP models the \textit{Quantum Router Software Architecture} (QRSA)~\cite{satoh2022quisp} shown in Fig.~\ref{fig:software_arch}(a).
At the heart of this software architecture is the notion of a \textit{RuleSet}~\cite{matsuo2019quantum}, which consists of a list of \textit{Condition} and \textit{Action Clauses}.
The actions are triggered upon satisfaction of local conditions.
For example, an entanglement swapping operation is executed on the stationary qubits at a node if the node shares entangled pairs with two of its neighbors.
The structure of RuleSets is not fixed but can be extended and defined by the user to achieve the desired node behavior~\cite{satoh2023rula}.

RuleSets are generated by the \textit{Connection Manager} and executed by the \textit{Rule Engine}.
Information about physical qubits inside the QNICs is gathered by the \textit{Hardware Monitor} and passed to the Connection Manager and the \textit{Routing Daemon}.

QuISP supports a number of error models for the stationary and flying qubits.
Qubits can undergo general Pauli noise, relaxation/excitation errors, and qubit loss.
Unitary operations applied during information processing by the quantum network nodes are not assumed to be perfect but are subject to gate errors.
In certain cases, such as for entanglement swapping, the nodes are required to measure some of their qubits.
Such measurements are subject to measurement errors.
Photon emission from a quantum memory and coupling to an optical fiber is also assumed to be probabilistic.
Finally, detectors used to measure flying qubits are characterized by their efficiency and dark count rates.
All of these error parameters can be defined by the user.

\subsection{SeQUeNCe} \label{sec:sequence}
Simulator of QUantum Network Communication (SeQUeNCe)~\cite{sequence} is an open-source discrete-event simulator that enables detailed, accurate, and highly customizable simulations of photonic quantum networks.
SeQUeNCe is developed in Python 3.12+ and it is designed to track billions of events per second of simulation time on a single machine~\cite{sequence-github}.
SeQUeNCe can be used to understand the trade-offs of alternative quantum network architectures, optimize quantum hardware, and study quantum networking protocols at scale. 

SeQUeNCe adopts a modularized design, shown in Fig.~\ref{fig:software_arch}(b), that enables researchers to customize the simulated network and reuse existing models in a flexible way.
The simulator is composed of five modules (hardware, entanglement management, resource management, network management, and applications) and a simulation kernel.
The simulation kernel provides an interface for discrete event simulation via a scheduler; it additionally tracks and updates the quantum state of simulation components via a quantum state manager.
The hardware module implements various quantum network components (e.g., quantum and classical channels, photon detectors and sources, and quantum memories), describing their behavior and modeling errors.
The entanglement management module includes protocols for reliable, high-fidelity distribution of entangled qubit pairs between end nodes in a quantum network. 
The resource and network management modules use classical control messages to manage allocation of local (per node) and network resources, respectively. 
Similar to QuISP, the resource manager uses an internal set of rules to manage the state of quantum memories.
The Application module represents quantum network applications and their service requests. 

One goal of SeQUeNCe is to provide realism in representing quantum states to accurately track quantum states in the network.
In SeQUeNCe, quantum states can be represented using the state vector, stabilizer~\cite{patange2026qre-cec}, density matrix, Bell-diagonal, or bosonic Fock state density matrix formalisms~\cite{zang2022simulation}, the latter three of which incorporate undetected errors in the states.
The quantum state manager provides external interfaces for generating simple operators, applying these operators to a quantum state, applying a generalized amplitude damping channel to a quantum state (modeling the decoherence that ruins a quantum state), or measuring a quantum state. For efficient simulation of imperfect bipartite entanglement distribution, SeQUeNCe also incorporates analytical error models assuming Pauli channels for initial entanglement generation, quantum memory decoherence, two-qubit gates, and single-qubit measurements~\cite{zang2026quantum}.
These allow users to construct hardware models with minimal knowledge of the underlying state representation.

\section{Methodology} \label{Benchmarking}
\begin{table*}[]
\centering
\caption{Summary of the experimental settings used in our cross-validation benchmarks. The first four rows list the metrics gathered in each experiment.}
\label{tab:experiments}
\resizebox{\textwidth}{!}{%
\begin{tabular}{l c c c c}
\toprule
 & \textbf{Exp 1} & \textbf{Exp 2} & \textbf{Exp 3} & \textbf{Exp 4} \\
 \midrule
 \textbf{Expected time} $T_{\text{exp}}(N_{\text{Bell}})$ & $N_{\text{Bell}} = 1000$ &$N_{\text{Bell}} = 1000$ & - & - \\
\textbf{Fidelity} $F$ & - & - & \checkmark & \checkmark \\
\textbf{Throughput} $N_{\text{pur}}$/sec & - & - & - & \checkmark \\
\textbf{Purification success probability} $p_{\text{succ}}$ & - & - & - & \checkmark \\
\midrule
\textbf{Distance between end nodes } $L$ & 20 km & 20 km & 40 km & 40 km\\
\textbf{BSA location } $d$ & 10 km & $\{10, 11, \ldots, 20\}$ km & 10 km & 20 km\\
\textbf{Memories per end node } $N_{\text{mem}}$ & $\{1, 2, 4, 8, 16\}$ & 1 & 1 & 2\\
\textbf{CNOT error prob. } $p_g$ & 0 & 0 & $0 \leq p_g \leq 1$ & 0 \\
\textbf{Measurement error prob. } $p_m$ & 0 & 0 & $0 \leq p_m \leq 1$ & 0 \\
\textbf{Memory coherence time } $\tau$ & $\infty$ & $\infty$ & $\{18$ ms, 55 ms, $\infty\}$ & $\{18$ ms, 55 ms, $\infty\}$ \\
\textbf{Initial Entanglement Fidelity} & 1 & 1 & 1 & $\{0.6, 0.65,\ldots, 0.9\}$ \\
\midrule
\textbf{Loss in fiber} & 0.2 dB/km & 0.2 dB/km & 0.2 dB/km & 0.2 dB/km \\
\textbf{Repetition rate} & 1 GHz & 1 GHz & 1 GHz & 1 GHz\\
\textbf{Speed of light in fiber } $c$ & $2 \times 10^5$ km/s & $2 \times 10^5$ km/s & $2 \times 10^5$ km/s & $2 \times 10^5$ km/s\\
\bottomrule
\end{tabular}%
}
\label{tab:experiments}
\end{table*}

Regardless of the internal workings of either of the two simulators, it is crucial that they both produce data reflecting physical reality.
When developing any simulator, one of the major issues is to build confidence in the accuracy of the simulation results.
A typical approach to the verification of simulators is to compare their predictions with analytical results~\cite{zang2024analytical}.
This is a natural approach, but due to the complexity of quantum networks, it can be done only for basic cases.
Furthermore, the design and implementation phase of simulator development may potentially introduce subtle assumptions about the theoretical model itself.
To uncover such hidden bias, one needs to either compare the simulator output with an actual experiment~\cite{Singal:24} or resort to cross-validation with another simulator. We cross-validate QuISP and SeQUeNCe through a set of carefully chosen experiments of increasing complexity to go beyond such limitations and increase confidence in simulation results. We focus on two main metrics in our cross-validation: time-to-generate and end-to-end fidelity. Time-to-generate is defined as the total time a simulator takes to satisfy a connection request. This is meant to cross-validate the probabilistic nature of entanglement generation in 1G repeater networks, as well as the implementation and timing regimes of classical messaging. The end-to-end entanglement fidelity aids in cross-validating the error models used in the simulators, to account for various sources of noise.

We propose four experiments targeting these two metrics under different scenarios.
The first two experiments focus on link-level entanglement generation by measuring the time required to generate $N_{\text{Bell}}$ base Bell pairs.
The third experiment investigates entanglement swapping and how the fidelity of the resulting end-to-end Bell pair is affected by different imperfections in the system. Finally, the fourth experiment studies entanglement purification, and how the time-to-generate and fidelity are affected by the differences in protocol design. A summary of the experiments can be found in Table~\ref{tab:experiments} along with the details of the simulation settings.

\textit{Experiment 1: Symmetric MIM Link.---} The first experiment tests the basic operation of an MIM link architecture with the BSA placed in the middle of the link, as shown in Fig.~\ref{fig:preliminaries}(a).
The total length of the link is $L=20$~km, meaning the BSA is placed $d=10$~km away from both nodes.
We vary the number of quantum memories, $N_{\text{mem}}$, inside the two quantum nodes.
The experiment is repeated for 1, 2, 4, 8, and 16 quantum memories.
We record the total time required to create $N_{\text{Bell}}=1000$ base Bell pairs.

\textit{Experiment 2: Asymmetric MIM link.---} This experiment also tests the operation of an MIM link but differs from Experiment 1 in the placement of the BSA, more realistically matching likely physical deployments.
Here, the BSA location is shifted gradually from the middle of the link towards one of the nodes in increments of 1~km until the link becomes a Memory-Memory (MM) link where the BSA can be thought of as part of the node.
The number of quantum memories is fixed to be $N_{\text{mem}}=1$, and we again measure the total time required to generate $N_{\text{Bell}}=1000$ base Bell pairs.

\textit{Experiment 3: Two-link entanglement swapping.---}
In this experiment, the two end nodes are separated by a distance of 40~km with a repeater node placed in the middle, 20~km away from each end-node.
The BSAs are placed halfway between the repeater and each end node, meaning $d=10$ km.
The repeater performs entanglement swapping on the two elementary link-level Bell pairs shared between each end-node and the repeater.
For this experiment, we measure the fidelity of the Bell pairs shared between the two end-nodes after successful entanglement swap.
Sources of noise that are varied are imperfect two-qubit CNOT gates, measurement errors, and decoherence of the quantum memories with three coherence times, $\{18~\text{ms}, 55~\text{ms}, \infty\}$.

\textit{Experiment 4: Entanglement purification.---}
Our final experiment focuses on entanglement purification~\cite{bennett1996mixed}, one of the basic forms of error management in quantum networks.
The experiment cross-validates simulations of one round of the BBPSSW protocol~\cite{bennett1996purification}, where two quantum network nodes first establish two Bell pairs, after which they execute bi-lateral CNOT gates using one Bell pair as the control and the other pair as the target.
Finally, the target pair is measured in the computational basis and the measurement results are exchanged between the nodes.
The protocol succeeds if the measurement outcomes agree.
We note that the full BBPSSW protocol includes an additional twirling step that we do not simulate. For this experiment, we focus on a single link between two nodes separated by 40~km, with the BSA placed in the middle (symmetric MIM link as in Experiment 1). We keep the number of memories per node fixed at $N_{\text{mem}}=2$. Similarly to Experiment 3, we consider three coherence times for the memories $\{18~\text{ms}, 55~\text{ms}, \infty\}$. We also compare the output of the  simulators on Bell pairs with different initial fidelities varying from $0.6$ to $0.9$ in increments of 0.05. The simulators are compared on the throughput of purified Bell pairs per second, the fidelity of the purified pairs, and the success probability of the purification process.

\section{Communication Models and Simulated Errors} \label{Models}
In this section, we discuss the different implementations of communication timing models in the two simulators, as well as their approach to modeling errors and noise.

\subsection{Communication Model for QuISP}\label{sec:comm-model-quisp}
\begin{figure*}
    \centering
    \includegraphics[width=\textwidth]{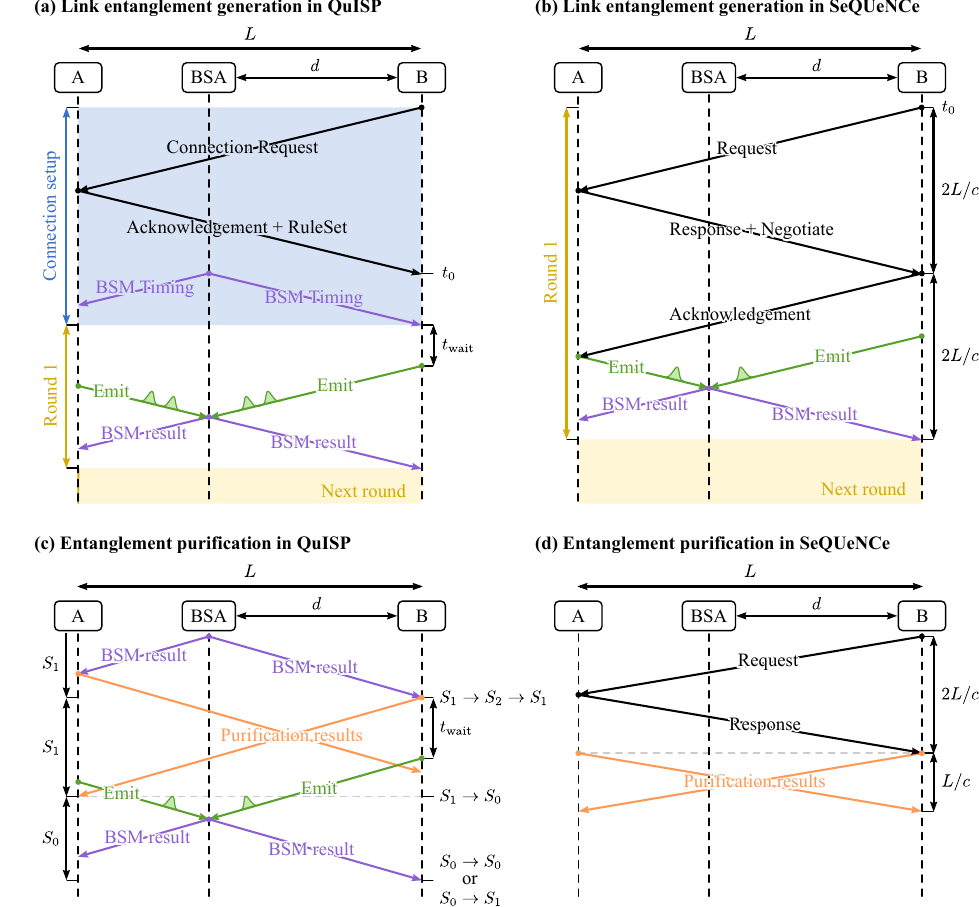}
    \caption{(a) Generation of link-level entanglement in QuISP. The photonic train is represented by a single green arrow. (b) Timing analysis for the entanglement generation protocol for SeQUeNCe. (c) Entanglement purification protocol in QuISP showing how the measured memories are freed even before the purification results reach both quantum nodes. (d) SeQUeNCe keeps the memories locked until the purification results are shared between the quantum nodes.}
\label{fig:sequance_diagram}
\end{figure*}

QuISP continuously generates link-level Bell pairs to fill up the quantum memories even if there are no connections that would utilize the resource.
To record the time for the experiments in Sec.~\ref{Benchmarking}, we create a connection between network nodes A and B following the process shown in Fig.~\ref{fig:sequance_diagram}(a).
The entanglement generation protocol begins with a connection setup phase, depicted inside the blue shaded area in Fig.~\ref{fig:sequance_diagram}(a).
Node B sends a connection request specifying the number of Bell pairs, $N_{\text{Bell}}$, that it would like to share with node A.
Node A replies with an acknowledgment message along with a generated RuleSet.
At this point, QuISP starts counting the time ($t=0$) that it takes to generate $N_{\text{Bell}}$ Bell pairs.
Also at $t=0$, the BSA sends a classical message instructing the quantum memories at nodes A and B when to emit photons in order for their arrival to be synchronized at the BSA for measurement, concluding the connection setup phase.

After the message from the BSA is received by the more distant node (in our case node B), the first round starts.
It consists of an initial wait period $t_{\text{wait}}$, during which the nodes prepare their memories for photon emission.
Each node emits a train of $N_{\text{mem}}$ photons towards the BSA.
After all photons are measured, provided they were not lost in transit, the BSA replies with the measurement results and new timings for the next emission of a photonic train.
This concludes one round, and the whole process is repeated if more Bell pairs are required.
The time stops when node B receives the notification from BSA that the final Bell pair was generated.

We can now quantify the expected time to generate $N_{\text{Bell}}$ pairs, denoted by $T_{\text{exp}}(N_{\text{Bell}})$.
The generated Bell pairs are immediately consumed, freeing their respective memories for further generation attempts.
We assume that the time for the BSA to perform its measurements and classical processing is negligible compared to signal propagation and other operations.

The time it takes for both nodes to receive a message from the BSA is given by $T_{\text{BSA}}=d/c$ given our assumption that $d\geq L/2$.
Here $c$ is the speed of light in optical fiber.
The initial two-way handshake of the connection setup phase takes $2L/c$, leading to the total time for the connection setup phase to be
\begin{equation}
    T_{\text{setup}} = (2L+d)/c.
\end{equation}
The photons in a train are separated by a time interval $t_{\text{sep}}$.
The value of this separation is governed by the physical extent of the wave-packet of individual photons, which in turn depends on the physical implementation of the quantum memories.
Parameter $t_{\text{sep}}$ ensures that the photons emitted from different memories of the quantum node do not overlap, which would otherwise lead to errors in tagging which photons underwent a successful BSM.

Before emitting the photons, the memories must spend time $t_{\text{wait}}$ to prepare for the emission. This time varies with the quantum technology used for the memories leading it to be also dependent on $t_{\text{sep}}$.
In QuISP, it is set to $t_{\text{wait}}=10t_{\text{sep}}$.
We can now write down an expression for the duration of one round,
\begin{equation}
    T_{\text{round}} = 2T_{\text{BSA}} + 10t_{\text{sep}} + (N_{\text{mem}}-1)t_{\text{sep}}.
\end{equation}

Finally, to estimate the total expected time to generate $N_{\text{Bell}}$ pairs, we first require the expected number of rounds,
\begin{equation}
    k = \left\lceil\frac{N_{\text{Bell}}}{N_{\text{mem}}p_{\text{link}}}\right\rceil,
    \label{eq:exp_number_rounds}
\end{equation}
where $p_{\text{link}}$ is the combined probability of both photons arriving at the BSA and then undergoing a successful BSM.
This leads to the expected time to generate $N_{\text{Bell}}$ Bell pairs,
\begin{equation}
    T(N_{\text{Bell}}) = T_{\text{setup}} + k T_{\text{round}}.
    \label{eq:time_quisp}
\end{equation}

\begin{figure}[t]
    \centering
    \includegraphics[width=\columnwidth]{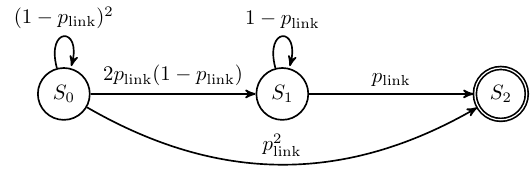}
    \caption{State transition diagram for generating two Bell pairs in QuISP.}
    \label{fig:markov-quisp}
\end{figure}

Equation~\eqref{eq:time_quisp} is enough to estimate the total time required for Experiments 1-2.
However, Experiment 4 on entanglement purification requires further discussion.
A crucial observation is that the memories become free at the end of each Round in Experiments 1-3.
Whether the BSM succeeds or fails, the memories are reinitialized for the next entanglement generation attempt in the following round.
In other words, the number of photons in the photon train is $N_{\text{mem}}$ in each Round.

This is not true for entanglement purification in Experiment 4, as depicted in
Fig.~\ref{fig:sequance_diagram}(c).
We denote by $S_i$, where $i=0,1,2$, the state of the memories, where $i$ represents the number of Bell pairs shared between the nodes.
For the purposes of illustrating the message exchange between the nodes in Fig.~\ref{fig:sequance_diagram}(c), we assume that the system starts in the state $S_1$, and that the purple BSM result message heralds successful generation of the second Bell pair. After this message reaches the further node (Node B in our case), the system transitions from $S_1\rightarrow S_2$.
Immediately, Node B performs its part of the purification protocol, transitioning the system from $S_2\rightarrow S_1$.
Regardless of whether the purification attempt succeeds, the measured qubits are immediately freed, wait for a time $t_{\text{wait}}$, and attempt to generate a new Bell pair, shown in green.
On the other hand, the unmeasured qubits must remain locked until the nodes exchange their purification measurement results, shown in orange.
Upon receiving these classical messages, the memories are freed and the system transitions from $S_1\rightarrow S_0$.
This means that the expected time to successfully obtain the first purified Bell pair differs from the expected time to obtain any of the remaining $N_{\text{pur}}-1$ purified Bell pairs.

Detailed derivation of the expected time $T_{\text{exp}}(N_{\text{pur}})$ is somewhat involved even for the case of $N_{\text{mem}}=2$, which is the case we assume here.
We model the process of generating two Bell pairs as a discrete-time Markov chain, shown in Fig.~\ref{fig:markov-quisp}, with $S_0$ being the initial and $S_2$ the accepting states.
The duration of each entanglement generation attempt, $t_{\text{link},i}$, depends on the memory state $S_i$,
\begin{equation}
    t_{\text{link},1} = t_{\text{wait}} + 2d/c, \quad t_{\text{link},0} = t_{\text{link,1}} + t_{\text{sep}}.
\end{equation}
We can now compute the expected time $T_i$ to reach state $S_2$ starting from state $S_i$, for $i=0,1$,
\begin{equation}
    T_1 = \frac{t_{\text{link},1}}{p_{\text{link}}}, \quad\text{and}\quad T_0 = \frac{t_{\text{link},0} + 2 (1 - p_{\text{link}}) t_{\text{link},1}}{p_{\text{link}} (2 - p_{\text{link}})}.
    \label{eq:pur-model-T1T0}
\end{equation}
Using these expected times to reach $S_2$, we can  obtain an expression for the expected time to reach $S_2$ following the first purification attempt, as well as all subsequent attempts,
\begin{equation}
    T_{\text{subsequent}} = t_{\text{link},1} + p_{\text{link}} \cdot T_1 + (1 - p_{\text{link}}) \cdot T_0.
\end{equation}
Finally, the expected time to deliver $N_{\text{pur}}$ purified Bell pairs is the sum of the expected time to succeed with purification for the first time and the expected time we succeed with the $N_{\text{pur}}-1$ subsequent attempts,
\begin{equation}
    T_{\text{exp}}(N_{\text{pur}}) = T_0 + \frac{N_{\text{pur}} - p_{\text{succ}}}{p_{\text{succ}}} \cdot T_{\text{subsequent}}.
    \label{eq:completion_time}
\end{equation}

\subsection{Communication Model for SeQUeNCe}
\label{sec:comm-model-sequence}
SeQUeNCe also continuously generates entanglement, assuming that an application consumes entanglement immediately upon generation. The communicating nodes A and B negotiate the time and availability of memories for generating entanglement, as shown in Fig.~\ref{fig:sequance_diagram}(b). The protocol is initiated ($t=0$) when a node designated as the primary (say, node B) sends a resource manager request to the other node (A) to generate entanglement. The protocol ends when the BSM node has communicated the end of the protocol to both the communicating parties. 
Unlike QuISP, SeQUeNCe restarts the generation protocol every time a new Bell pair is to be generated.

The structure of the entanglement generation protocol for SeQUeNCe makes timing analysis straightforward. Nodes A and B perform a three-way handshake to agree on the protocol parameters for entanglement generation, taking $3L/c$.
When multiple memories are present, the protocol is run for each memory pair simultaneously. The photon emissions are time multiplexed into bins which distinguish the transmitted photons according to the pairs of memories communicating with each other. We ignore the time delay introduced due to the time multiplexing when multiple memories operate simultaneously.
The time required for one round of entanglement generation using the three-way handshake is given by
\begin{equation}
    T_{\text{round}} = \frac{3L}{c} + \frac{L-d}{c}+\frac{d}{c} = \frac{4L}{c}
\end{equation}

Hence, to generate $N_{\text{Bell}}$ Bell pairs, SeQUeNCe is expected to take
\begin{equation}
    T_{\text{exp}}(N_{\text{Bell}}) = k(4L/c),
    \label{eq:time_sequence}
\end{equation}
where $k$ is the expected number of rounds defined in~\eqref{eq:exp_number_rounds}.
Due to the fact \sequence performs the connection setup phase for every Bell pair, we expect the total time to be longer than \quisp's for Experiments 1-3 which do not include purification. For Experiment 4, which includes purification, the total time estimation for \sequence must be updated similar to \quisp.

The \sequence purification process begins after the successful establishment of two Bell pairs, as in \quisp. However, establishing a set of two Bell pairs for purification is performed through an additional negotiation round between nodes A and B, as shown in Fig.~\ref{fig:sequance_diagram}(d) as black arrows. After a successful negotiation, one of the two Bell pairs is measured, and the measurement result is communicated according to the orange arrows. Neither the measured nor unmeasured memories are freed for further entanglement generation attempts until after both nodes receive the purification result. The negotiation round along with the heralding messages ensure that the entanglement purification attempts will never overlap in timing with generation attempts, unlike the model present in \quisp.

The total time for one purification attempt in \sequence is thus given from Fig.~\ref{fig:sequance_diagram}(d) as $t_{\mathrm{EPP}} = 3L/c$.

The separation in timing for purification and generation allows us to calculate explicitly the expected delivery time and output fidelity for $N$ purified Bell pairs, which we derive in Appendix~\ref{sec:sequence-epp}. To summarize, we first calculate the expected time to generate two Bell pairs using two memories,
\begin{equation}
    T_{\textrm{2pair}} = \frac{3-2p_\mathrm{link}}{p_\mathrm{link}(2-p_\mathrm{link})}\frac{4L}{c},
\end{equation}
with the success probability of entanglement generation $p_\mathrm{link}$ as defined previously. The total expected time for one round of two-pair generation and a purification attempt is then $T_\mathrm{2pair} + 3L/c$ considering the additional time for purification. With a complete generation and success probability $p_\mathrm{tot}$, we then have that
\begin{equation}
    T_\mathrm{exp}(N_\mathrm{pur}) = \frac{N}{p_\mathrm{tot}}\left(\frac{3-2p_\mathrm{link}}{p_\mathrm{link}(2-p_\mathrm{link})}\frac{4L}{c} + \frac{3L}{c}\right).
    \label{eq:purification-time-sequence}
\end{equation}
Finally, considering the decoherence models in \sequence described in Sec.~\ref{sec:sequence-error}, the probability $p_\mathrm{tot}$ may be analytically solved with initial entanglement fidelity (before decoherence) $F$ as
\begin{gather}
    p_\mathrm{tot} = \frac{(1-p_\mathrm{link})\left[9(e^{\frac{8L}{c\tau}} - 1) + 2p_\mathrm{link}(8F_0^2 - 4F_0 + 5)\right]}{9(2-p_\mathrm{link})(e^{\frac{8L}{c\tau}} - 1 + p_\mathrm{link})}\nonumber\\
    + \frac{p_\mathrm{link}(8F_0^2 -4F_0 +5)}{9(2-p_\mathrm{link})},
    \label{eq:sequence-p-tot} \\
    F_0 = F e^{-4L/c\tau} + \frac{1-e^{-4L/c\tau}}{4}, \label{eq:fid-decohere}
\end{gather}
which we derive in Appendix~\ref{sec:sequence-epp} using the decoherence model in Section~\ref{sec:sequence-error}. We note the additional decoherence imposed in~\eqref{eq:fid-decohere} arises from the round-trip negotiation in the \sequence purification protocol implementation.

\subsection{Error models for SeQUeNCe}\label{sec:sequence-error}
\begin{figure}[t]
    \centering
    \includegraphics[width=\columnwidth]{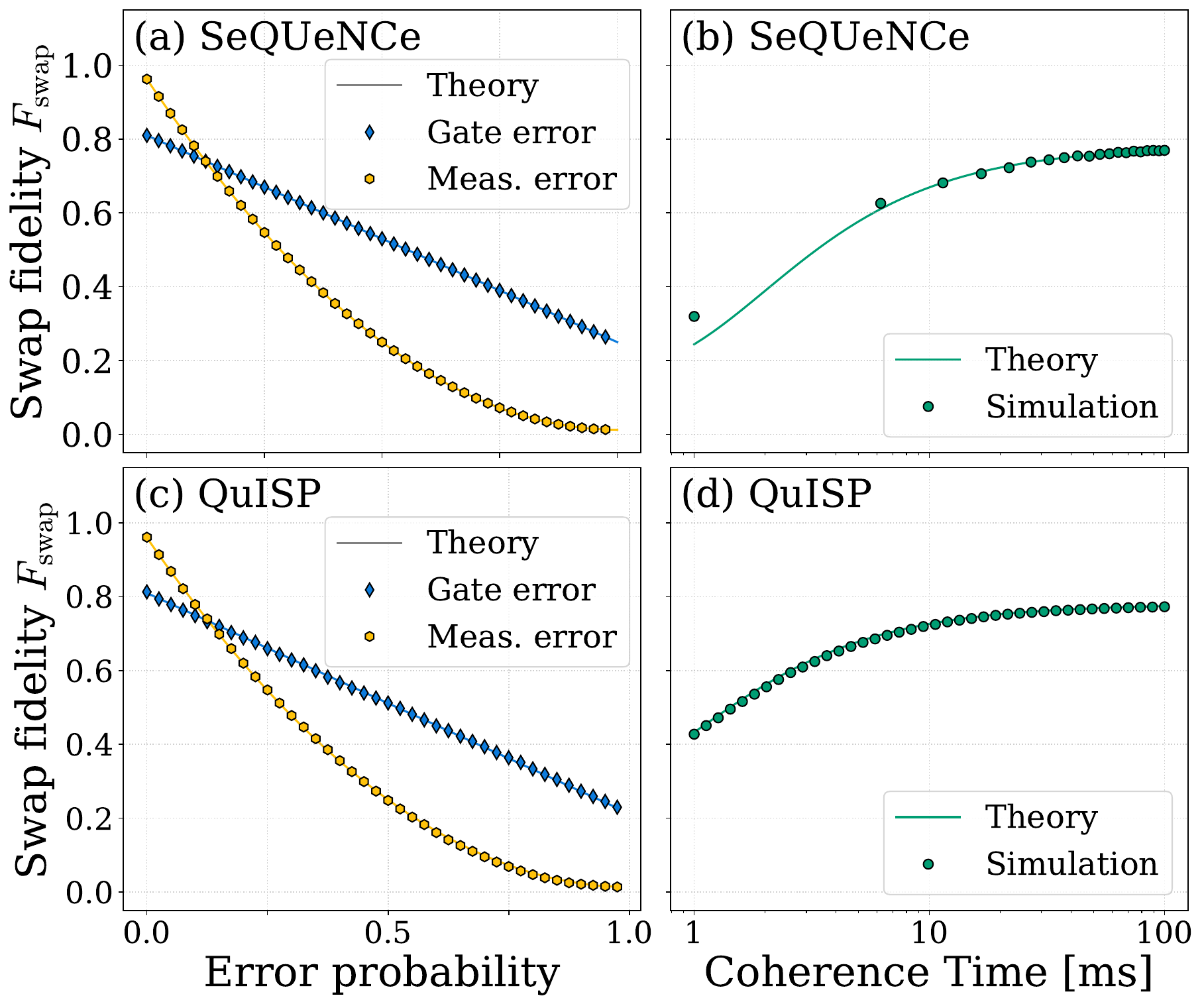}
    \caption{(a) Verification of the \sequence error model without decoherence in Eq.~(\ref{eq:sequence-fidelity-verification}). The measurement error $p_{\text{m}}=0.1$ is shown in the yellow points, and the gate error $p_{\text{g}}=0.05$ is shown in blue with initial Werner state fidelities, $F_1, F_2 = 1$.
    (b) Verification of Eq.~(\ref{eq:sequence-fidelity-verification}) for SeQUeNCe when $p_{\text{m}}=0.1$ and $p_{\text{g}}=0.05$, varying the degree of depolarizing noise affecting the memories.
    (c) Verification Eq.~(\ref{eq:quisp-swap-fidelity-perfect-memory}), for the end-to-end fidelity of QuISP. The blue diamonds correspond to the fidelity when $p_{\text{g}}$ varies and $p_{\text{m}}=0.1$, while the yellow hexagons correspond to varying $p_{\text{m}}$ with $p_{\text{g}}=0.05$. Both simulators show good agreement between the obtained data and their respective theoretical models shown in solid lines.}
    \label{fig:error_models}
\end{figure}

Here we specify the error models used by SeQUeNCe for quantum processes which happen in the quantum networking scenarios under our investigation. Specifically, quantum states which are stored in quantum memories decohere over time, and the quantum gates and measurements used in entanglement protocols, especially entanglement swapping for this work, are imperfect. We will use the \textit{fidelity} $F$ to quantify the probability that what we believe the state to be corresponds to the true state (i.e., no error has occurred).

For quantum memory decoherence, SeQUeNCe implements the analytical model of continuous-time Pauli channel for light-weight evaluation of memory errors during quantum network processes. For the cross-validation experiments, we focus on the depolarizing channel where the occurrences of three different Pauli errors, $X$ (bit flip), $Z$ (phase flip) and $Y$ (both), are equally likely; a state that has passed through such a channel is known as a \textit{Werner state}. The effect on a single qubit is to preserve the quantum state with probability $p$ and to bring the state into a maximally mixed state with probability $1-p$. We can model the time-dependence of the depolarizing channel with $p=\exp(-t/\tau)$, where $\tau$ is the memory coherence time. 

If a Werner state of fidelity $F_{\text{in}}$ is stored in two quantum memories with identical coherence times $\tau$ for time $t$, the final state is still a Werner state but with a degraded fidelity~\cite{zang2023entanglement,zang2025entanglement}:
\begin{align}
    F(t) = F_{\text{in}} e^{-2t/\tau} + \frac{1-e^{-2t/\tau}}{4}.
    \label{eq:sequence-decohere}
\end{align}

For imperfect quantum operations, especially two-qubit gates and single-qubit measurements, SeQUeNCe implements the following analytical models also for the ease of tracking operational errors. For any two-qubit gate, with probability $1-p_{\text{g}}$ the expected gate operation is applied noiselessly and with probability $p_{\text{g}}$ the two involved qubits are completely depolarized (i.e., the state has become pure noise).

For single-qubit measurements in the computational basis,  $1-p_{\text{m}}$ is the probability for the measurement to give the correct outcome, and with probability $p_{\text{m}}$ the classical measurement outcome is flipped.

Under the above operational error models, now we consider entanglement swapping based on noisy CNOT gate and imperfect single-qubit measurement, with two Werner states as input. The output fidelity can be analytically expressed as~\cite{zang2023entanglement}:
\begin{align}
    F_\mathrm{swap,W} & = \frac{p_{\text{g}}}{4} + (1-p_{\text{g}})\left[(1-p_{\text{m}})^2\left(F_1F_2 + 3e_1e_2\right)\right.\nonumber \\
    & \left. +p_{\text{m}}(2 - p_{\text{m}})\left(F_1e_2 + e_1F_2 + 2e_1e_2\right)\right],
    \label{eq:sequence-fidelity-verification}
\end{align}
where $F_1,F_2$ are the fidelities of the input Werner states, and we have defined $e_i=(1-F_i)/3$ to simplify the notation.

We verify our swapping model for imperfect operations in Fig.~\ref{fig:error_models}(a).
We observe excellent agreement between~\eqref{eq:sequence-fidelity-verification} and the output of \sequence when the two-qubit gates or the measurements are noisy.
Effect of decoherence in the form of depolarizing noise is depicted in Fig.~\ref{fig:error_models}(b).
Again, we observe good agreement between our model and simulator.

For the BBPSSW purification protocol in Experiment 4, the output fidelity of a successful entanglement purification round may be analytically derived for two known input fidelities $F_1$ and $F_2$ as~\cite{bennett1996purification,deutsch1996quantum}
\begin{gather}
    F_\mathrm{succ}(F_1,F_2) = \frac{F_1F_2 + \frac{(1-F_1)(1-F_2)}{9}}{p_\mathrm{succ}},
\end{gather}
where $p_\mathrm{succ}$ is the probability of purification success and we assume no gate or measurement errors as in Table~\ref{tab:experiments}. For this work, the link success rate $p_\mathrm{link}$ defined for the communication models allows us calculate a distribution of possible decoherence times for one memory while awaiting the second successful entanglement generation, as described in Section~\ref{sec:comm-model-sequence}. This distribution then leads to a distribution of input fidelities according to the decoherence model in~\eqref{eq:sequence-decohere}, which may be used to find the average purification output fidelity and success probability. The expected output fidelity of shown in~\eqref{eq:purification-fidelity-sequence}, and is derived in  Appendix~\ref{sec:sequence-epp}.
\begin{figure*}[h]
\normalsize
\begin{equation}
\label{eq:purification-fidelity-sequence}
F'= \left[\frac{(1-p_\mathrm{link})[3(e^{\frac{8L}{c\tau}} - 1)(2F_0 + 1) + 4p_\mathrm{link}(10F_0^2 - 2F_0 +1)]}{18(2-p_\mathrm{link})(e^{\frac{8L}{c\tau}} - 1 + p_\mathrm{link})} + \frac{p_\mathrm{link}[9F_0^2 + (1-F_0)^2]}{9(2-p_\mathrm{link})}\right] / p_\mathrm{tot}.
\end{equation}
\hrulefill
\vspace*{4pt}
\end{figure*}
The probability of purification success weighted by expected purification time $p_\mathrm{tot}$ is defined in~\eqref{eq:sequence-p-tot} and the round-trip decohered fidelity $F_0$ is defined in~\eqref{eq:fid-decohere}.
Figure~\ref{fig:exp4_quisp}(a) shows the verification of SeQUeNCe's timing and noise models for delivering 500 purified Bell pairs using the BBPSSW purification protocol and memories with 55~ms coherence time.

\subsection{Error Models for QuISP}

QuISP treats time-dependent errors, such as depolarization and photon loss, by discrete exponentiation of a time-invariant state transition matrix.
QuISP associates each qubit with an \textit{error vector probability},
\begin{equation}
    \vec{\pi}(t) = (\pi_I, \pi_X, \pi_Y, \pi_Z, \pi_R, \pi_E, \pi_L),
\end{equation}
where $\pi_j$ represents the probability of the qubit being in the state affected by error $j$ at time $t$, and $\sum_j \pi_j = 1$.
Errors $\{X,Y,Z\}$ represent Pauli errors, $\{R,E\}$ are relaxation and excitation errors, and $L$ is photon loss. Index $I$ denotes no error and no loss.

The error probability vector evolution is described by a \textit{transition matrix} $Q$, where each entry $Q_{ij}$ represents the probability of transitioning from the error state $\pi_i$ to state $\pi_j$ per time $\delta t$.
Each row of $Q$ preserves the probability and thus satisfies $\sum_j Q_{ij} = 1$.
The error vector probability at time $t$ is then given by
\begin{equation}
    \vec{\pi}(t) = \vec{\pi}(0) Q^t.
\end{equation}
QuISP allows users to specify error parameters by setting values of elements $Q_{ij}$.
Before performing any quantum operation, \quisp needs to sample the error probability vector turning the qubits into a definite pure state known to the simulator but not to the network protocols and applications being modeled.
For the single-qubit measurement error, \quisp performs a perfect measurement on the qubit, and with error probability $p_m$ flips the reported measurement outcome.

For two-qubit gates, \quisp first applies the noiseless gate followed by applying a gate error with probability $p_g$. 
If the gate is determined to be noisy, a two-qubit Pauli operator is sampled from the set of 15 possible errors $(\{I, X, Y, Z\} \otimes \{I, X, Y, Z\} - \{I\otimes I\})$ with user specified weights (for depolarizing error, they are all equal).
Assuming that link-level Bell pairs are noiseless, the fidelity of the end-to-end Bell pair is
\begin{align}
    F_\mathrm{swap} & = (1 - p_{\text{g}})  (1 - p_{\text{m}})^2 + \frac{3}{15} p_{\text{g}}  (1 - p_{\text{m}})^2 \nonumber\\
        &+ \frac{8}{15} p_{\text{g}}  (1 - p_{\text{m}}) p_{\text{m}} + \frac{4}{15} p_{\text{g}}  p^2_{\text{m}}.
\label{eq:quisp-swap-fidelity-perfect-memory}
\end{align}

To account for memory decoherence, \quisp models the noise on all memories as identical depolarizing channels governed by the transition matrix $Q$.
Because Bell states are maximally entangled and symmetric, applying identical depolarizing channels to distributed qubits is mathematically equivalent to applying the combined time-evolution of that noise to just one qubit.
Therefore, we can aggregate the decoherence effects from all involved memories into a single fidelity term,
\begin{equation}
    F_\text{mem} = \left(Q^{2t_1 + 2t_2 + 2T}\right)_{00},
\end{equation}
where $t_1$ ($t_2$) is the time elapsed from photon emission until the link-level Bell pair is created for node A (B), and $T$ is the propagation time for the swap measurement results to reach the end nodes from the repeater.

Because both the swap operation and the memory decoherence act as depolarizing channels on the state, we calculate the final end-to-end fidelity by composing these two channels as
\begin{equation}
    F_{\text{total}} = F_{\text{mem}} F_{\text{swap}} + \frac{\left(1-F_{\text{mem}}\right)\left(1-F_{\text{swap}}\right)}{3}.
\label{eq:quisp-swap-fidelity-with-decoherence}
\end{equation}

For depolarizing errors, we have a closed-form expression for any integer power of the transition matrix,
\begin{equation}
    \left(Q^t_\mathrm{depol}\right)_{00} = \frac{1+3^{1-t}(3-4 p)^t}{4},
\end{equation}
where $p$ is chosen such that $\left(Q^{t_\mathrm{coherence}}_\mathrm{depol}\right)_{00} = 1/e$.

We confirm that our predicted end-to-end fidelities accurately describe QuISP's output in the absence of photon loss and for perfect BSM with $p_{\text{BSA}}=1$.
Figure~\ref{fig:error_models}(c) shows the end-to-end fidelity as a function of varying either the gate error or the measurement error while keeping the other at a fixed value.
Our predictions using~\eqref{eq:quisp-swap-fidelity-perfect-memory} display excellent agreement with \quisp.
In Fig.~\ref{fig:error_models}(d), we include the effect of depolarizing noise while fixing both gate and measurement errors to a finite value and again observe agreement between~\eqref{eq:quisp-swap-fidelity-with-decoherence} and the output of the simulator.
We should note that the two-qubit gate error probability and the coherence time definitions between SeQUeNCe and QuISP are not the same.
To convert to and from SeQUeNCe convention to QuISP, we need the following equations,
\begin{align}
    T^\mathrm{QuISP}_\mathrm{coherence} &= T^\mathrm{SeQUeNCe}_\mathrm{coherence} \log\left(\frac{3 e}{4 - e}\right), \\
    p^\mathrm{QuISP}_\mathrm{CNOT} &= \frac{15}{16} p^\mathrm{SeQUeNCe}_\mathrm{CNOT}.
\end{align}

The purification protocol in Experiment~4 can be viewed as a four-qubit $ZZZZ$ stabilizer measurement, meaning that after a successful purification the two Bell pairs are projected onto their +1 eigenspace.
Given that we focus on depolarizing noise only and that the two-qubit noise model is equivalent to a model affecting only one of the qubits at twice the error probability, the $i$-th Bell pair can be described by a simplified error vector
\begin{equation}
    \mathbf{e}^{(i)}=(\pi^{(i)}_I,\pi^{(i)}_X,\pi^{(i)}_Y,\pi^{(i)}_Z)^T.
\end{equation}
The probability of purification success given two noisy Bell states $\mathbf{e}^{(1)}$ and $\mathbf{e}^{(2)}$ is given by
\begin{align}
    p_{\text{succ}} & = (\pi_I^{(1)}+\pi_Z^{(1)})(\pi_I^{(2)}+\pi_Z^{(2)}) \nonumber\\
    & + (\pi_X^{(1)}+\pi_Y^{(1)})(\pi_X^{(2)}+\pi_Y^{(2)}).
    \label{eq:success_prob}
\end{align}
Given a successful purification, the state of the unmeasured Bell pair is described by the following error vector,
\begin{equation}
    \mathbf{e}^{(1)}_{\text{out}} = \frac{1}{p_{\text{succ}}} \begin{pmatrix}
        \pi_I^{(1)}\pi_I^{(2)} + \pi_Z^{(1)}\pi_Z^{(2)} \\
        \pi_X^{(1)}\pi_X^{(2)} + \pi_Y^{(1)}\pi_Y^{(2)} \\
        \pi_X^{(1)}\pi_Y^{(2)} + \pi_Y^{(1)}\pi_X^{(2)} \\
        \pi_I^{(1)}\pi_Z^{(2)} + \pi_Z^{(1)}\pi_I^{(2)}
    \end{pmatrix}.
    \label{eq:output_purification}
\end{equation}

\begin{figure}[t]
    \centering
    \includegraphics[width=\columnwidth]{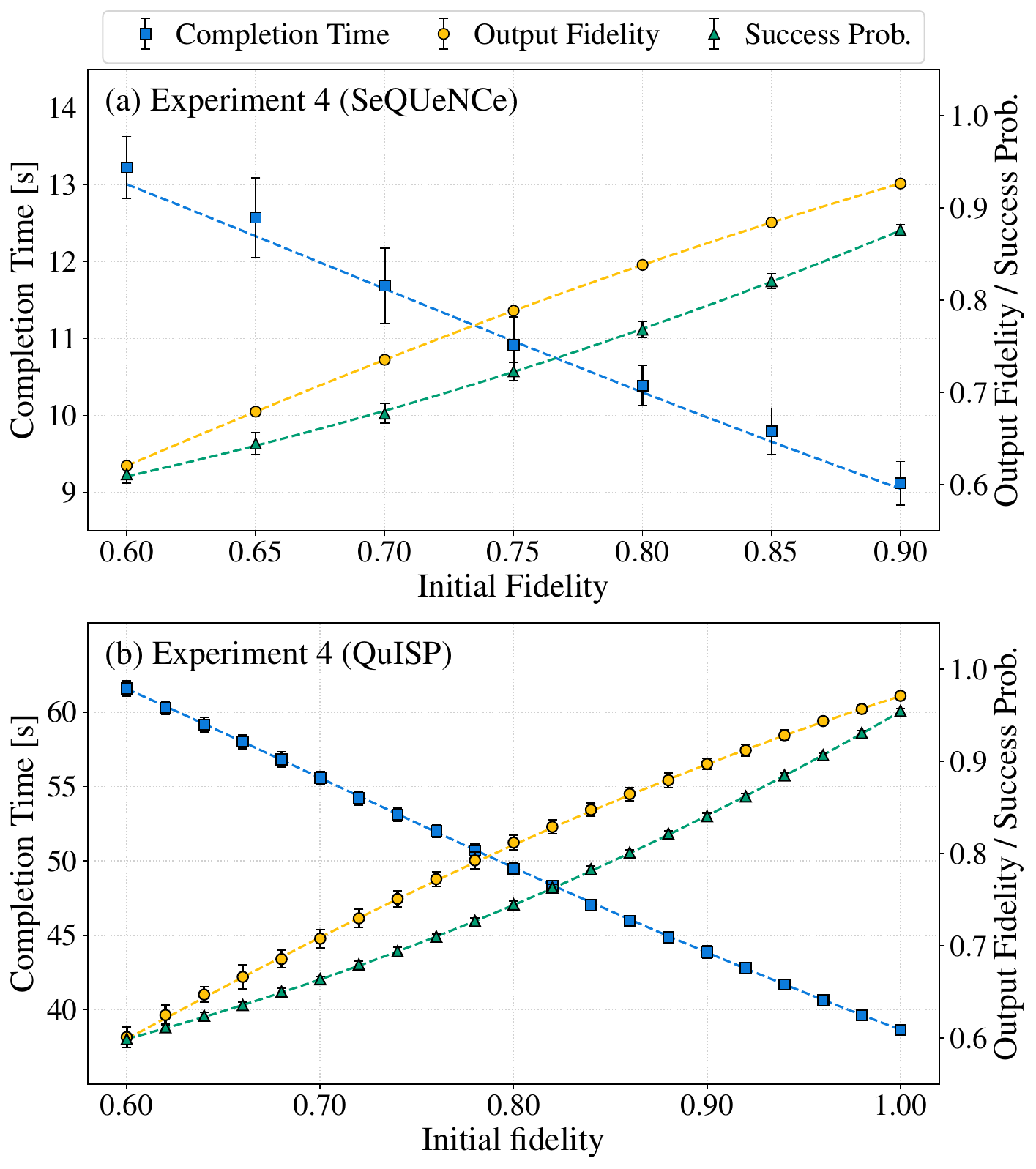}
    \caption{Verification of the timing and noise models for X-purification protocol and 55~ms coherence time for the memories. (a) For SeQUeNCe, the expected completion time for $N_{\text{pur}=500}$ is given by~\eqref{eq:purification-time-sequence}, the success probability is given by~\eqref{eq:sequence-p-tot}, and the final fidelity of a purified Bell pair by~\eqref{eq:purification-fidelity-sequence}. (b) For QuISP, the theoretical completion time is given by~\eqref{eq:completion_time}, where $N_{\text{pur}}=10^4$. The output fidelity of the purified Bell pairs and the success probability of purification are given by~\eqref{eq:output_purification} and~\eqref{eq:success_prob}, respectively.}
    \label{fig:exp4_quisp}
\end{figure}

We can now verify the simulation results of the X-purification protocol in QuISP against its theoretical model.
Figure~\ref{fig:exp4_quisp}(b) shows the completion time to deliver $10^4$ purified Bell pairs, the output of these pairs as well as the success probability of the purification protocol.
The simulation results for all three metrics agree very well with the corresponding theoretical models.

\section{Results} \label{Results}
\subsection{Symmetric MIM link} \label{sec:symmetric-MIM-link}

\begin{figure}[t]
    \centering
    \includegraphics[width=\columnwidth]{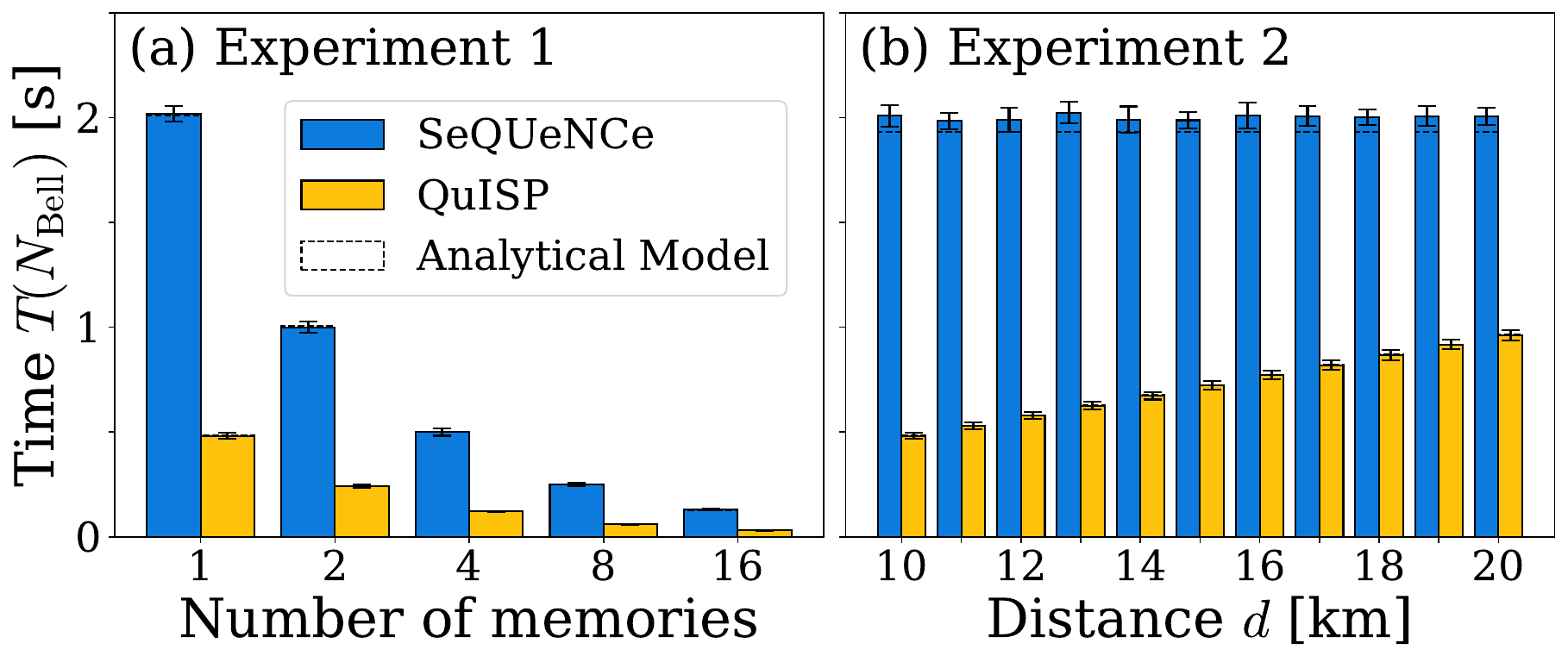}
    \caption{(a) Time to generate 1000 Bell pairs for a symmetric MIM link for Experiment 1. Generation time is shown for both SeQUeNCe and QuISP using 1, 2, 4, 8, and 16 memories at each node. (b) Time to generate 1000 Bell pairs for an asymmetric MIM link for Experiment 2. Generation time is shown for SeQUeNCe and QUISP for increasing asymmetries between a node and the BSM.}
    \label{fig:exp1_2}
\end{figure}
The results of Experiment~1 are shown in Fig.~\ref{fig:exp1_2}(a), where the total time to generate 1000 Bell pairs is plotted against the number of quantum memories available at each node.
The dashed line represents theoretic predictions of~\eqref{eq:time_quisp} for QuISP and~\eqref{eq:time_sequence} for SeQUeNCe.
Both simulators agree excellently with their respective theoretical models, and also with each other on the general qualitative behavior that utilizing multiple memories leads to faster distribution of the requested number of Bell pairs.
However, the simulators disagree on the actual time it takes to do so, with SeQUeNCe being slower than QuISP.

This is explained by the two different approaches that the simulators take when setting up the connection between A and B.
SeQUeNCe performs a three-way handshake for every single Bell pair as explained in Fig.~\ref{fig:sequance_diagram}(b) while QuISP performs a two-way handshake only once.
Given that this is the main difference, one would expect the ratio of the two times to be roughly constant for varying number of memories.
This is indeed the case, with the ratio varying only slightly between 4.16 and 4.33. 

\subsection{Asymmetric MIM Link}\label{sec:asymetric-MIM-link}
The case of the asymmetric MIM link pictured in Fig.~\ref{fig:exp1_2}(b) shows qualitative difference between the simulators.
The time it takes to setup a connection in SeQUeNCe is independent of the actual placement of the BSA.
This is evident also from~\eqref{eq:time_sequence} for the theoretical expected time $T(N_{\text{Bell}})$. 
QuISP's connection setup protocol on the other hand performs best when the BSA is located in the middle, and its performance deteriorates as the distance between the BSA and node B increases.

\subsection{Two-link entanglement swapping}
\begin{figure}
    \centering
    \includegraphics[width=\columnwidth]{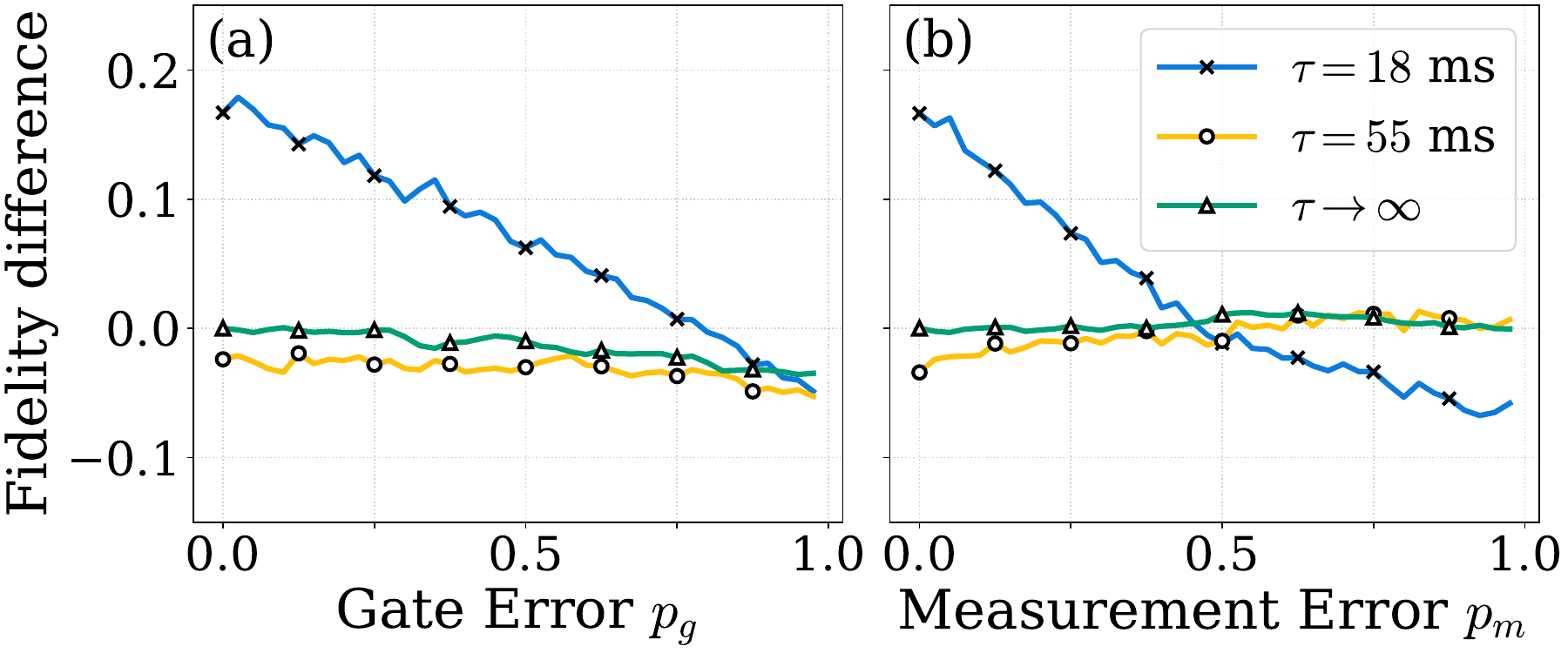}
    \caption{Cross-validation of QuISP and SeQUeNCe in Experiment 3 for three  memory coherence times. Normalized fidelity difference $\Delta F_{\text{swap}}$ from~\eqref{eq:fidelity_diff_swap} of the end-to-end entangled pair is shown as a function of the gate error $p_g$ in (a), and as a function of the measurement error $p_m$ in (b).}
    \label{fig:exp3_cross}
\end{figure}

We now discuss the results of the two-link entanglement swapping experiment under the influence of noisy gates, measurements, and depolarizing quantum memories.
The metric used to compare the two simulators is the normalized relative swapping fidelity difference,
\begin{equation}
    \Delta F_{\text{swap}} = \frac{F^{Q}_{\text{swap}} - F^{S}_{\text{swap}}}{F^{S}_{\text{swap}}},
    \label{eq:fidelity_diff_swap}
\end{equation}
where $F^{Q}_{\text{swap}}$ and $F^{S}_{\text{swap}}$ are the swapping fidelities obtained from \quisp and \sequence, respectively.
Fig.~\ref{fig:exp3_cross}(a) shows $\Delta F_{\text{swap}}$ as a function of the CNOT gate error $p_g$ for $18$ ms, $55$ ms, and infinite coherence times.
We observe general agreement between the two simulators for the $55$ ms and infinite coherence times. In Fig.~\ref{fig:exp3_cross}(b) we show $\Delta F_{\text{swap}}$ as a function of the measurement error $p_m$ for the the same coherence times.

The simulators disagree on the quantitative behavior of end-to-end fidelity for quantum memories with poor coherence times, with \quisp predicting higher values for low error probability range while predicting lower values for higher error probability range.
This can be traced back to the discussed differences in the simulators' communication models and their slight difference in the two-qubit gate error models. 
\quisp is generally faster when it comes to generation of link-level entanglement, as seen in Fig.~\ref{fig:exp1_2}(a).
This means the link-level Bell pairs spend less time in the memories before undergoing entanglement swapping. Higher base fidelities subsequently lead to higher end-to-end quality of the distributed Bell pairs.

\subsection{Symmetric MIM Link Entanglement Purification}
\begin{figure}[t]
    \centering
    \includegraphics[width=\columnwidth]{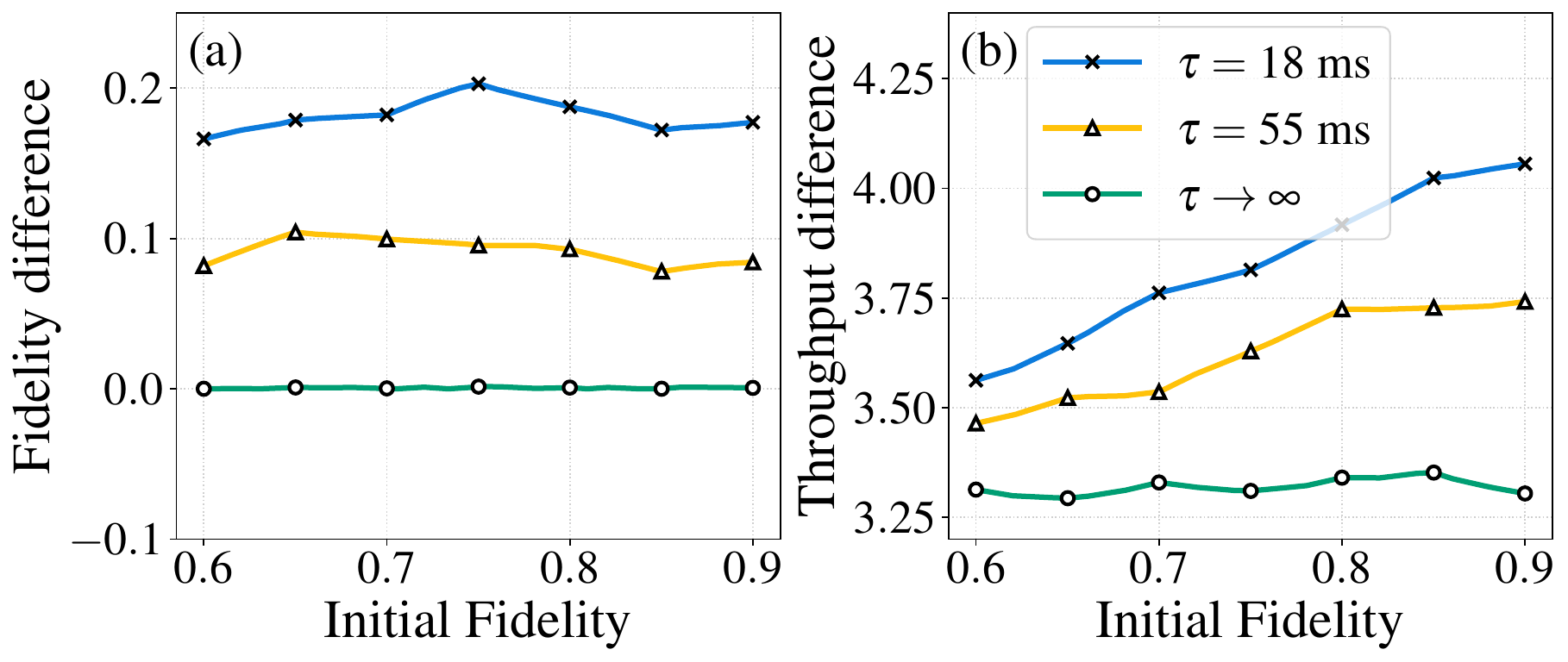}
    \caption{Cross-validation of the two simulators using Experiment 4. (a) Normalized fidelity difference $\Delta F_{\text{pur}}$ of the successfully purified Bell pairs. (b) Normalized throughput difference, measured in Bell pairs per second.}
    \label{fig:exp4_cross}
\end{figure}

In Fig.~\ref{fig:exp4_cross}, we compare the results for link purification Experiment 4 under depolarizing emissive memories.
The fidelity difference in Fig.~\ref{fig:exp4_cross}(a) is defined similarly to the swapping fidelity difference of~\eqref{eq:fidelity_diff_swap},
\begin{equation}
    \Delta F_{\text{pur}} = \frac{F^{Q}_{\text{pur}} - F^{S}_{\text{pur}}}{F^{S}_{\text{pur}}}.
    \label{eq:fidelity_diff_pur}
\end{equation}
We observe that for perfect memories ($\tau\rightarrow\infty$), the two simulators agree perfectly.
For finite memory coherence times, the fidelity difference increases but remains nearly constant across the range of initial Bell pair fidelities.

The observed independence of $\Delta F_{\text{pur}}$ from the initial fidelity is expected.
The output fidelity of a successfully purified Bell pair does depend on the initial fidelities of the two link Bell pairs as seen in~\eqref{eq:purification-fidelity-sequence} and~\eqref{eq:output_purification}.
However, the average time the first generated link Bell pair has to wait for the second one is independent of the initial fidelity.
Therefore, the decrease in output fidelity is equal for both simulators, resulting in constant fidelity difference for all initial fidelities.
The difference between the two simulators becomes apparent for memories with finite coherence time.
Owing to different communication models, the time the first link Bell pair idles before purification takes place varies between the two simulators.
As discussed in Sec.~\ref{Models}, QuISP distributes link Bell pairs faster, leading to shorter idling time for the first pair.
This in turn results in higher output fidelity of the purified Bell pair for a given initial fidelity compared to SeQUeNCe.

The effects of the two communication models can be observed on the throughput difference in Fig.~\ref{fig:exp4_cross}(b) as well.
Throughput is defined as the number of successfully purified Bell pairs that are delivered to the application per second.
It is both a function of how quickly the link Bell pairs can be generated and the success probability of purification, which in turn depends on the fidelities of the link Bell pairs.
For perfect memories ($\tau\rightarrow\infty$), the success probability of purification is solely determined by the initial fidelity of the Bell pairs, so that only the speed of the communication model determines the throughput.
This results in a constant throughput difference, with QuISP being slightly faster as expected.
The situation changes for memories with finite coherence times, where the slower communication model of SeQUeNCe has two consequences.
It directly affects the throughput by generating Bell pairs a slower rate, while also keeping the Bell pairs idling for longer, which affects the success probability of purification.
This compounds into a linear dependence of the throughput difference on the initial fidelity as seen in Fig.~\ref{fig:exp4_cross}(b).

\section{Discussion} \label{Discussion}
\label{sec:discussion}
We have performed a cross-validation of \quisp and \sequence, two prominent open-source simulators for quantum networks.
We studied the protocols that both simulators use to generate link-level entanglement in order to evaluate how long each simulator takes to satisfy a connection request.
\quisp relies on a protocol requiring fewer classical messages, leading to faster rate of link-level entanglement compared to \sequence.
Interestingly, we discovered that the slower connection setup protocol for \sequence is insensitive to the placement of the BSA node, a property that might be of interest when planning deployment of first generation quantum networks.

We also focused on how the simulators handle noisy operations and decoherence of quantum memories in both entanglement swapping and purification scenarios.
The simulators agreed on the general qualitative behavior but differed on the specific values of end-to-end fidelity.
This can be understood by the different communication models utilized by both simulators. The interplay between differences in both communication models and error models will be explored in more detail as future work.

The tests presented here have strengthened our confidence in both simulators and deepened our understanding of the performance impact of the differences in the modeled protocols.  The tests described in this paper are deliberately chosen to also be analytically tractable, allowing us to write down closed-form equations, which match the outcomes of the simulations.  With these results in hand, the obvious next step is to move beyond analytically tractable cases and compare the results of larger and more complex systems.
Examples of such systems include longer chains of quantum repeaters with complex entanglement swapping and purification policies, different link architectures, and different quantum repeater generations. Our cross-validation experiments also pave the way to benchmarking quantum networks with non-trivial topologies as well as multiplexed networks servicing multiple independent connection requests simultaneously.

Despite the existence of suites of software quality assurance (SQA) tests for both simulators, this exercise has not only given us a deeper understanding of the network design choices but also resulted in numerous outright bug fixes to both simulators.  We recommend this direct comparison approach for both quantum and classical network simulators where possible.

\section*{Code Availability}
Source code, documentation, and sample
configurations are fully available online at \cite{quisp-github} and \cite{sequence-github} for both QuISP and SeQUeNCe, respectively.
Data of the simulation results and configuration files for the experiments will be made available upon publication of this manuscript.

\clearpage  

\begin{appendices}

\section{Analytical modeling for entanglement purification scenario in SeQUeNCe}\label{sec:sequence-epp}
Here we describe the scenario of entanglement purification protocol (EPP) in the quantum repeater which we use for cross-validation in this work, and demonstrate the analytical modeling of this scenario with stochastic analysis~\cite{zang2024analytical}. 

\subsection{Setup}
\label{sec:sequence-epp-setup}
The setup and assumptions are described as follows.
\begin{enumerate}
    \item Consider a single link between two directly connected quantum repeater nodes. Each node only has 2 quantum memories, so at most 2 entangled pairs can be shared between the two nodes at any specific time point. 
    \item Entanglement will be generated probabilistically between each quantum memory pair in an independent way. Each attempt requires time $T$, and succeeds with probability $p$.
    \item The raw entangled state, i.e. the entangled state between two quantum memories immediately after successful generation, is in the form of Werner state, with initial fidelity $F$. More specifically, we consider that the raw state is the state at the end of the success entanglement generation attempt, i.e. after the success heralding signal has been received by both nodes.
    \item Each quantum memory decoheres after being initialized, e.g. after starting to store entangled state, and the decoherence is described by single-qubit depolarizing channel with decoherence time $\tau$.
    \item The canonical bilocal CNOT 2-to-1 EPP will be applied immediately after two entangle pairs are available~\cite{zang2023entanglement,zang2025entanglement} for higher success probability. We assume that both nodes will receive the measurement results $t_\mathrm{EPP}$ time after the EPP circuit is run.
\end{enumerate}

\subsection{Stochastic model}
The probability for each pair of quantum memories to get entangled after $n$ attempts (in terms of time, after $nT$ time) is
\begin{align}
    \mathrm{Pr}_\mathrm{A/B}(n) = (1-p)^{n-1}p,
\end{align}
where we use A and B to label two pairs of quantum memories. The quantum memories will not be initialized after entanglement is successfully generated, so at the first time when both entangled pairs have been successfully created, the later generated entangled pair at time $t$ is in its raw form, while the other pair generated at $t'\leq t$ has decohered for $\Delta t = t'-t$ time. In the following, we denote the entangled pairs by the time they have decohered for. The input states of the EPP will be $\rho_A(0)$ and $\rho_B(\Delta t)$, where we have assumed that A is generated after B without loss of generality. Then the successful output state of the EPP is $\rho_\mathrm{out} = \mathcal{E}[\rho_A(0),\rho_B(\Delta t)]$, where $\mathcal{E}[\cdot,\cdot]$ denote the successful output state for general input pair. The success probabibility of the EPP is $p_\mathcal{E}[\rho_A(0),\rho_B(\Delta t)]$, where $p_\mathcal{E}[\cdot,\cdot]$ denote the success probability for general input pair. 

\subsubsection{Probability distribution of successful entanglement generation time interval}
The success probability of the EPP and the successful output state of the EPP only depend on the difference between the times when the two entangled pairs are generated. Therefore, now we consider the probability distribution of $\Delta t$. Note that in the assumed scenario time is always discrete with unit being $T$. The probability for pair A to successfully generate entanglement at $\delta T$ ($\delta = 0,1,2,\dots$) before pair B successfully generates entanglement is
\begin{align}
    \mathrm{Pr}[n_\mathrm{B} - n_\mathrm{A} = \delta] =& \sum_{n=1}^\infty \mathrm{Pr}_\mathrm{A}(n) \mathrm{Pr}_\mathrm{B}(n + \delta)\nonumber\\
    =& \sum_{n=1}^\infty \left[(1-p)^{n-1}p\right] \left[(1-p)^{n+\delta-1}p\right]\nonumber\\
    =& \frac{p}{2-p} (1-p)^\delta.
\end{align}
By symmetry, it has the same probability for pair A to successfully generate entanglement at $\delta T$ after pair B successfully generates entanglement. When we focus on the output of the EPP, we only care about the absolute interval between the times when two entangled pairs are generated. Therefore, we have
\begin{align}
    & \mathrm{Pr}[|n_\mathrm{B} - n_\mathrm{A}| = \delta] = \frac{2p}{2-p} (1-p)^\delta, \\
    & \mathrm{Pr}[n_\mathrm{B} = n_\mathrm{A}] = \frac{p}{2-p}.
\end{align}
To be more concise in the following we will use $p_\mathrm{interval}(\delta)$ to denote $\mathrm{Pr}[|n_\mathrm{B} - n_\mathrm{A}| = \delta]$ with $\delta\geq 0$.

\subsubsection{Probability distribution of successful output fidelity}
With the probability for the successful entanglement generation time interval, we can obtain the statistics of successful EPP output fidelity, and EPP success probability, accumulating results from each attempt of entanglement generation + EPP. For the specific error models listed above (can be generalized to other initial state and memory decoherence model), we have the Bell fidelity after both quantum memories have decohered for time $t$ (note the transform between different parameterizations)~\cite{zang2025entanglement}
\begin{align}\label{eqn:depolarizing}
    F(t) = F(0)e^{-2t/\tau} + \frac{1 - e^{-2t/\tau}}{4}.
\end{align}
The fidelity of the EPP successful output state when the two input states are Werner states with fidelity $F_1$ and $F_2$ is~\cite{bennett1996purification,deutsch1996quantum}
\begin{align}
    &F_\mathrm{succ}(F_1,F_2)\nonumber\\
    =& \frac{F_1F_2 + \frac{(1-F_1)(1-F_2)}{9}}{F_1F_2 + \frac{F_1(1-F_2)+(1-F_1)F_2}{3} + \frac{5(1-F_1)(1-F_2)}{9}},
\end{align}
which is the fidelity of $\rho_\mathrm{out}(0)$. As the success of EPP can only be confirmed after both nodes receive the communication of the measurement results, the real final fidelity will be $\rho_\mathrm{out}(t_\mathrm{EPP})$. The success probability of the EPP when the two input states are Werner states with fidelity $F_1$ and $F_2$ is~\cite{bennett1996purification,deutsch1996quantum}
\begin{align}
    p_\mathrm{succ}(F_1,F_2) =& F_1F_2 + \frac{F_1(1-F_2)+(1-F_1)F_2}{3}\nonumber\\
    &+ \frac{5(1-F_1)(1-F_2)}{9},
\end{align}
so that $p_\mathcal{E}[\rho_A(0),\rho_B(\delta T)] = p_\mathrm{succ}[F(0),F(\delta T)]$. Then the probability for the output state to have fidelity $F_\mathrm{succ}[F(0),F(\delta T)]$, conditioned on successful EPP is
\begin{align}
    p_\mathrm{fid-succ}(\delta) \coloneqq& \mathrm{Pr}[F(\rho_\mathrm{out}) = F_\mathrm{succ}[F(0),F(\delta T)]] \nonumber\\
    =& \frac{p_\mathrm{succ}[F(0),F(\delta T)] p_\mathrm{interval}(\delta)}{\sum_{\delta=0}^\infty p_\mathrm{succ}[F(0),F(\delta T)] p_\mathrm{interval}(\delta)}.
\end{align}
We have the closed-form expressions for the denominator which is the total success probability (for each complete attempt of entanglement generation and entanglement purification)
\begin{align}\label{eqn:succ_fid}
    p_\mathrm{tot-succ} \coloneqq& \sum_{\delta=0}^\infty p_\mathrm{succ}[F(0),F(\delta T)] p_\mathrm{interval}(\delta)\nonumber\\
    =& p_\mathrm{succ}[F(0),F(0)] p_\mathrm{interval}(0)\nonumber\\
    &+ \sum_{\delta=1}^\infty p_\mathrm{succ}[F(0),F(\delta T)] p_\mathrm{interval}(\delta)\nonumber\\
    =& \frac{(1-p)\left[9(e^{\frac{2T}{\tau}} - 1) + 2p(8F^2 - 4F + 5)\right]}{9(2-p)(e^{\frac{2T}{\tau}} - 1 + p)}\nonumber\\
    &+ \frac{p(8F^2 -4F +5)}{9(2-p)}.
\end{align}
Then we have the closed form of expected output fidelity conditioned on successful EPP
\begin{align}\label{eqn:output_fid}
    \bar{F}' =& \sum_{\delta=0}^\infty F_\mathrm{succ}[F(0),F(\delta T)] p_\mathrm{fid-succ}(\delta)\nonumber\\
    =& F_\mathrm{succ}[F(0),F(0)] p_\mathrm{fid-succ}(0)\nonumber\\
    &+ \sum_{\delta=1}^\infty F_\mathrm{succ}[F(0),F(\delta T)] p_\mathrm{fid-succ}(\delta)\nonumber\\
    =& \left[\frac{(1-p)[3(e^{\frac{2T}{\tau}} - 1)(2F + 1) + 4p(10F^2 - 2F +1)]}{18(2-p)(e^{\frac{2T}{\tau}} - 1 + p)}\right.\nonumber\\
    &+ \left. \frac{p[9F^2 + (1-F)^2]}{9(2-p)}\right] / p_\mathrm{tot-succ}.
\end{align}
As visualized in the main text, the expected successful output fidelity is not guaranteed to improve over the initial fidelity, which is an instantiation of the non-universality of entanglement purification protocols~\cite{zang2025no} in realistic quantum networking scenarios.

\subsubsection{Probability distribution of each complete generation and purification cycle duration}
In the previous section we have ignored the duration it takes each attempt of entanglement generation plus entanglement purification. Here we examine the timing effect of this single-link scenario.

In the current setup, a complete generation and purification cycle will end after an EPP attempt, with success leading to the increment of the entangled state counter, and failure leading to retry. Therefore, we must wait till both quantum memory pairs have successfully generated entanglement so that we can attempt EPP. The probability for memory pair A to successfully generate entanglement at time $n_AT>0$, and memory pair B to successfully generate entanglement at time $n_BT>0$ is
\begin{align}
    p_\mathrm{time}(n_A,n_B) \coloneqq& \mathrm{Pr}[t_A=n_AT \land t_B=n_BT]\nonumber\\
    =& p^2(1-p)^{n_A+n_B-2}.
\end{align}
On the other hand, the memory decoherence and thus the success probability of EPP depends on the difference between $t_A$ and $t_B$, i.e. $|n_A-n_B|$. By symmetry, we may separate the cases where $n_A=n_B$ and the cases where $n_A\neq n_B$, and for the latter we can assume that $n_A>n_B$ without loss of generality, while noting that it only counts half the possibilities when $n_A\neq n_B$. For each cycle of entanglement generation and purification, suppose the cycle succeeds after $nT$ time for entanglement generation, we have its probability for $n>1$
\begin{align}
    &p_\mathrm{cycle-succ}(n)\nonumber\\
    =& 2 \sum_{\delta = 1}^{n-1}p^2(1-p)^{2n - \delta - 2} p_\mathrm{succ}[F(0),F(\delta T)]\nonumber\\
    &+ p^2(1-p)^{2n - 2} p_\mathrm{succ}[F(0),F(0)]\nonumber\\
    =& \frac{p(1-p)^{n-1}e^{\frac{2T}{\tau}}}{9\left[(1-p)e^{\frac{2T}{\tau}} - 1\right]}\biggl[9 (1-p)\left[1 - (1-p)^{n-1}\right] \nonumber\\
    & + e^{-\frac{2T}{\tau}}\left[(1-p)^{n-1} ((4F-1)^2p + 9) - 9\right] - p(4F-1)^2 \biggr]\nonumber\\
    &+ \frac{1}{9}(1-p)^{2n-2}p^2(8F^2 - 4F + 5),
\end{align}
while $p_\mathrm{cycle-succ}(1) = p^2p_\mathrm{succ}[F(0),F(0)]$. One can verify that $\sum_{n=1}^{\infty} p_\mathrm{cycle-succ}(n) = p_\mathrm{tot-succ}$, which further validates the derivations. The total success probability for each complete generation and purification cycle is still $p_\mathrm{tot-succ}$. 

Now we consider the time it takes to successfully purify $N$ entangled pairs. Let the $i$-th complete cycle take $t_i$ time, and in total it take $i_N$ complete cycles to purify $N$ entangled pairs. Then the total time is
\begin{align}
    t_N = \sum_{i=1}^{i_N}t_i.
\end{align}
The expected time to purify $N$ entangled pairs is the expectation value of the above random variable
\begin{align}
    \overline{t_N} = \mathbb{E}\left[\sum_{i=1}^{i_N}t_i\right] = \mathbb{E}[i_N]\mathbb{E}[t_i],
\end{align}
where for the second equality we have used the Wald's identity. We have from negative binomial distribution that
\begin{align}
    \mathbb{E}[i_N] = \frac{N}{p_\mathrm{tot-succ}}.
\end{align}
The probability for the complete cycle to take $nT$ for entanglement generation attempts is
\begin{align}
    p_\mathrm{duration}(n) =& p^2(1-p)^{2n - 2} + 2 \sum_{\delta = 1}^{n-1}p^2(1-p)^{2n - \delta - 2}\nonumber\\
    =& [2 - (2 - p)(1 - p)^{n-1}] (1 - p)^{n-1} p,
\end{align}
for $n>1$, and $p_\mathrm{duration}(1) = p^2$.
The expected time for entanglement generation attempts within each complete cycle (no matter success or failure) is
\begin{align}
    \mathbb{E}[t_i] =& p^2T + T\sum_{n=2}^{\infty}n\cdot p_\mathrm{duration}(n)\nonumber\\
    =& \frac{3 - 2p}{p(2-p)}T.
\end{align}
Note that for each complete cycle we may further add the fixed time overhead from the entanglement purification attempt $t_\mathrm{EPP}$. Combining all the above, we have the expected time for purifying $N$ entangled pairs
\begin{align}\label{eq:expected-time}
    \overline{t_N} = \frac{N}{p_\mathrm{tot-succ}} \left(\frac{3 - 2p}{p(2-p)}T + t_\mathrm{EPP}\right).
\end{align}

\subsubsection{Modification assuming additional time delay}
As detailed in main text section \ref{sec:comm-model-sequence}, the implementation of entanglement purification in \sequence incurs an additional two round-trip time delay before running the purification circuit. The purpose of this additional delay is to add an additional negotiation round and coordinate purification efforts between the two nodes. However, this additional delay violates assumption (5) in section~\ref{sec:sequence-epp-setup} above, which requires modification to the expected purification success probability, output fidelity, and time for purification from equations \ref{eqn:succ_fid}, \ref{eqn:output_fid}, and \ref{eq:expected-time}, respectively. Equations for deriving the entanglement generation expected time are unaffected by this implementation change.

To adjust equations \ref{eqn:succ_fid} and \ref{eqn:output_fid}, we note that each memory involved in purification will undergo decoherence for exactly one additional round-trip time $t_\mathrm{RT}$ before purification is attempted. The fidelities in the equations may thus be accurately stated by adjusting the fidelity after generation $F$. Concretely, we may rewrite equations \ref{eqn:succ_fid} and \ref{eqn:output_fid} by substituting $F$ with the decohered
\begin{equation}
    F_0 = F e^{-2t_\mathrm{RT}/\tau} + \frac{1-e^{-2t_\mathrm{RT}/\tau}}{4}
\end{equation}
where we use the sequence decoherence model from main text equation~\ref{eq:sequence-decohere}.

The modification of \ref{eq:expected-time} is even more straightforward. As we have already adjusted $p_\mathrm{tot-succ}$, and the first term of the equation only involves this quantity and quantities derived from the generation process, it requires no further modification. We only need to increase the time $t_\mathrm{EPP}$ to include the additional round-trip delay $t_\mathrm{RT}$ in the purification process.

\end{appendices}

\bibliographystyle{IEEEtran.bst}
\bibliography{IEEEabrv, bibfile}
\EOD
\end{document}